\documentclass{imsart}
\usepackage{natbib}
\usepackage{amsthm}
\usepackage{amsmath}
\usepackage{amsfonts}
\usepackage{bm}

\usepackage{algpseudocode}
\usepackage{algorithm}
\usepackage{booktabs}
\usepackage{tabularx}
\usepackage[draft]{todonotes}
\usepackage[colorlinks,citecolor=blue,urlcolor=blue,filecolor=blue,backref=page]{hyperref}
\usepackage{graphicx}

\startlocaldefs
\endlocaldefs

\begin{document}

%% *** Frontmatter *** 

\begin{frontmatter}
\title{Bayesian exponential random graph models for populations of networks}

%\title{\thanksref{T1}}
%\thankstext{T1}{<thanks text>}
\runtitle{}

\begin{aug}
%\author{\fnms{} \snm{}}
\author{\fnms{Brieuc} \snm{Lehmann}\thanksref{addr1,addr2} \ead[label=e1]{brieuc.lehmann@bdi.ox.ac.uk}}
\and
\author{\fnms{Simon} \snm{White}\thanksref{addr3, addr4}\ead[label=e2]{simon.white@mrc-bsu.cam.ac.uk}}

\runauthor{}

\address[addr1]{
Department of Statistics, University of Oxford, 29 St Giles', Oxford, OX1 3LB, United Kingdom
 }
 
\address[addr2]{
Big Data Institute, University of Oxford, Old Road Campus, Oxford, OX3 7LF, United Kingdom,
 \printead{e1} 
 }

\address[addr3]{
Department of Psychiatry, University of Cambridge, Cambridge, CB2 OAH, United Kingdom,
 \printead{e2} 
 }

\address[addr4]{
MRC Biostatistics Unit, University of Cambridge, Cambridge, CB2 0SR, United Kingdom
 }

%\thankstext{<id>}{<text>}

\end{aug}

\begin{abstract}

The collection of data on populations of networks is becoming increasingly common, where each data point can be seen as a realisation of a network-valued random variable. A canonical example is that of brain networks: a typical neuroimaging study collects one or more brain scans across multiple individuals, each of which can be modelled as a network with nodes corresponding to distinct brain regions and edges corresponding to structural or functional connections between these regions. Most statistical network models, however, were originally proposed to describe a single underlying relational structure, although recent years have seen a drive to extend these models to populations of networks. Here, we propose one such extension: a multilevel framework for populations of networks based on exponential random graph models. By pooling information across the individual networks, this framework provides a principled approach to characterise the relational structure for an entire population. To perform inference, we devise a novel exchange-within-Gibbs MCMC algorithm that generates samples from the doubly-intractable posterior. To illustrate our framework, we use it to assess group-level variations in networks derived from fMRI scans, enabling the inference of age-related differences in the topological structure of the brain's functional connectivity.

\end{abstract}

%% ** Keywords **

\end{frontmatter}

%% ** Mainmatter **

\section{Introduction}\label{background}

The statistical analysis of network data is becoming increasingly commonplace, with applications across various disciplines, such as epidemiology, social science, neuroscience and finance \citep{Kolaczyk2009}. Over the last four decades, a number of statistical models for networks have been developed, including stochastic blockmodels \citep{Holland1983}, latent space models \citep{Hoff2002}, and the focus of this article, exponential random graph models (ERGMs; \cite{Frank1986}). 

An exponential random graph model is a set of parametric statistical distributions on network data (see \cite{Schweinberger2020} for a recent review). The aim of the model is to characterise the distribution of a network in terms of a set of \textit{summary statistics}. These summary statistics are typically comprised of topological features of the network, such as the number of edges and subgraph counts. The summary statistics enter the likelihood via a weighted sum; the weights are (unknown) model parameters that quantify the relative influence of the corresponding summary statistic on the overall network structure and must be inferred from the data. ERGMs are thus a flexible way in which to describe the global network structure as a function of network summary statistics. 

To date, statistical network models, including ERGMs, have largely focused on the analysis of a single network. Formally, a network consists of a set of nodes and a set of edges between these nodes. Let $\mathcal{N} = \lbrace 1, \dots, N \rbrace $ be a finite set of nodes, each of which may be associated with covariates $x_i \in \mathcal{X} \subseteq \mathbb{R}^q$. An edge from node $i$ to node $j$ is denoted by $Y_{ij}$, so that the network is encoded by the adjacency matrix $\pmb{Y} = (Y_{ij})_{i,j \in \mathcal{N}}$. For our purposes, the set of nodes $\mathcal{N}$ and their covariates $\pmb{x} = \lbrace x_1, \dots, x_N \rbrace$ are considered fixed, while the edges are considered to be random variables. Denote $\pmb{y}$ to be an instantiation, or outcome, of the random adjacency matrix $\pmb{Y}$ and write $\mathbb{P}(\pmb{Y} = \pmb{y}) := \pi(\pmb{y})$ for the probability that $\pmb{Y}$ takes the value $\pmb{y}$. A statistical network model specifies a parametrised probability distribution on the adjacency matrix $\pi(\pmb{y} | \pmb{x}, \theta)$ where $\theta$ is a vector of model parameters. 

A population of networks consists of $n > 1$ adjacency matrices $\pmb{Y}^{(1)}, \dots, \pmb{Y}^{(n)}$ defined on a common set of nodes $\mathcal{N}$. We will assume for simplicity that the nodal covariates are the same across networks, though in principle this not need be the case. A common example of a population of networks arises in neuroimaging, where a typical study consists of brain data across a number of participants, each constituting an individual network. Network analyses of the brain can provide insight into cognitive function by revealing how distinct brain areas work in conjunction \citep{Fuster2006}. These analyses aim to identify salient topological features of the brain's connectivity structure that are common across individuals or that differ across groups.

While one could fit a single model to each individual network separately, it is not straightforward to combine these individual results into a single coherent result that is representative of the whole population. An alternative approach is to construct a group-representative network by, for example, taking the mean of the edges across the individual networks and applying a threshold to the resulting weighted network \citep{Achard2006}. These approaches ignore the individual variability present in the networks and, moreover, typically do not accurately summarise the topological information across the individual networks \citep{Ginestet2011}.

A more statistical approach is to treat each individual networks as distinct statistical units arising from a joint probability distribution $\pi(\pmb{y}^{(1)}, \dots,  \pmb{y}^{(n)}| \pmb{x}, \theta)$ \citep{Ginestet2017}. Here, we propose a multilevel framework for populations of networks based on Bayesian exponential random graph models. By pooling information across the individual networks, this framework provides a principled approach to characterise the relational structure for an entire population. Our method can be used to infer group-level differences in the network structure between \textit{sets of networks}, which we demonstrate on both simulated networks and real networks derived from resting-state functional magnetic resonance imaging (fMRI) scans from an ageing study.

Inference for Bayesian ERGMs is challenging due to the double-intractability of the ERGM posterior distribution; standard MCMC schemes such as the Metropolis algorithm are not feasible as it is not possible to evaluate the acceptance ratio. A common approach is to apply the exchange algorithm \citep{Murray2006}, which was first employed in the context of Bayesian ERGMs by \cite{Caimo2011}. To perform inference for our framework for populations of networks, we propose an \textit{exchange-within-Gibbs} algorithm that combines the exchange algorithm with the Gibbs sampler \citep{Geman1984} to produce samples from the target posterior distribution. The parameterisation of general multilevel models can play an important role in the overall efficiency of an MCMC scheme \citep{Gelfand1995, Papaspiliopoulos2003, Papaspiliopoulos2007}. To improve the mixing properties of the algorithm, we use an ancillarity-sufficiency interweaving strategy (ASIS) that interweaves between the \textit{centered} and non-centered parameterisations \cite{Yu2011}. To further boost efficiency, we also employ adaptation of the random-walk proposal parameters in the algorithm (see e.g. \cite{roberts1997weak}).

\subsection{Related work}

\subsection*{Hierarchical ERGMs}

\begin{figure}
    \centering
\begin{tikzpicture}[x=0.75pt,y=0.75pt,yscale=-1,xscale=1]
%uncomment if require: \path (0,414); %set diagram left start at 0, and has height of 414

%Rounded Rect [id:dp872848334583724] 
\draw   (30,65.52) .. controls (30,40.38) and (50.38,20) .. (75.52,20) -- (259.99,20) .. controls (285.13,20) and (305.51,40.38) .. (305.51,65.52) -- (305.51,202.1) .. controls (305.51,227.24) and (285.13,247.62) .. (259.99,247.62) -- (75.52,247.62) .. controls (50.38,247.62) and (30,227.24) .. (30,202.1) -- cycle ;
%Rounded Rect [id:dp2827435494599291] 
\draw  [dash pattern={on 0.84pt off 2.51pt}] (61.51,53.82) .. controls (61.51,44.88) and (68.77,37.62) .. (77.71,37.62) -- (127.31,37.62) .. controls (136.26,37.62) and (143.51,44.88) .. (143.51,53.82) -- (143.51,102.42) .. controls (143.51,111.37) and (136.26,118.62) .. (127.31,118.62) -- (77.71,118.62) .. controls (68.77,118.62) and (61.51,111.37) .. (61.51,102.42) -- cycle ;
%Rounded Rect [id:dp7823540269973116] 
\draw   (24.51,313.02) .. controls (24.51,303.96) and (31.86,296.62) .. (40.91,296.62) -- (97.11,296.62) .. controls (106.17,296.62) and (113.51,303.96) .. (113.51,313.02) -- (113.51,362.22) .. controls (113.51,371.28) and (106.17,378.62) .. (97.11,378.62) -- (40.91,378.62) .. controls (31.86,378.62) and (24.51,371.28) .. (24.51,362.22) -- cycle ;
%Shape: Ellipse [id:dp39505435349162754] 
\draw  [fill={rgb, 255:red, 0; green, 0; blue, 0 }  ,fill opacity=1 ] (99.05,51.7) .. controls (99.05,49.01) and (101.37,46.84) .. (104.23,46.84) .. controls (107.09,46.84) and (109.41,49.01) .. (109.41,51.7) .. controls (109.41,54.38) and (107.09,56.56) .. (104.23,56.56) .. controls (101.37,56.56) and (99.05,54.38) .. (99.05,51.7) -- cycle ;
%Shape: Ellipse [id:dp6973010119942644] 
\draw  [fill={rgb, 255:red, 0; green, 0; blue, 0 }  ,fill opacity=1 ] (72.29,72.76) .. controls (72.29,70.07) and (74.61,67.9) .. (77.47,67.9) .. controls (80.33,67.9) and (82.65,70.07) .. (82.65,72.76) .. controls (82.65,75.44) and (80.33,77.62) .. (77.47,77.62) .. controls (74.61,77.62) and (72.29,75.44) .. (72.29,72.76) -- cycle ;
%Shape: Ellipse [id:dp15498968055319085] 
\draw  [fill={rgb, 255:red, 0; green, 0; blue, 0 }  ,fill opacity=1 ] (82.65,104.35) .. controls (82.65,101.66) and (84.97,99.49) .. (87.83,99.49) .. controls (90.69,99.49) and (93.01,101.66) .. (93.01,104.35) .. controls (93.01,107.03) and (90.69,109.21) .. (87.83,109.21) .. controls (84.97,109.21) and (82.65,107.03) .. (82.65,104.35) -- cycle ;
%Shape: Ellipse [id:dp9121505114619302] 
\draw  [fill={rgb, 255:red, 0; green, 0; blue, 0 }  ,fill opacity=1 ] (123.22,72.76) .. controls (123.22,70.07) and (125.54,67.9) .. (128.4,67.9) .. controls (131.26,67.9) and (133.58,70.07) .. (133.58,72.76) .. controls (133.58,75.44) and (131.26,77.62) .. (128.4,77.62) .. controls (125.54,77.62) and (123.22,75.44) .. (123.22,72.76) -- cycle ;
%Shape: Ellipse [id:dp8517442879248023] 
\draw  [fill={rgb, 255:red, 0; green, 0; blue, 0 }  ,fill opacity=1 ] (115.45,103.54) .. controls (115.45,100.85) and (117.77,98.68) .. (120.63,98.68) .. controls (123.49,98.68) and (125.81,100.85) .. (125.81,103.54) .. controls (125.81,106.22) and (123.49,108.4) .. (120.63,108.4) .. controls (117.77,108.4) and (115.45,106.22) .. (115.45,103.54) -- cycle ;
%Shape: Ellipse [id:dp3235643471823404] 
\draw  [fill={rgb, 255:red, 0; green, 0; blue, 0 }  ,fill opacity=1 ] (229.19,105.92) .. controls (229.19,103.27) and (231.5,101.12) .. (234.37,101.12) .. controls (237.23,101.12) and (239.54,103.27) .. (239.54,105.92) .. controls (239.54,108.58) and (237.23,110.72) .. (234.37,110.72) .. controls (231.5,110.72) and (229.19,108.58) .. (229.19,105.92) -- cycle ;
%Shape: Ellipse [id:dp9074227867740788] 
\draw  [fill={rgb, 255:red, 0; green, 0; blue, 0 }  ,fill opacity=1 ] (202.43,76.32) .. controls (202.43,73.67) and (204.75,71.52) .. (207.61,71.52) .. controls (210.47,71.52) and (212.79,73.67) .. (212.79,76.32) .. controls (212.79,78.98) and (210.47,81.12) .. (207.61,81.12) .. controls (204.75,81.12) and (202.43,78.98) .. (202.43,76.32) -- cycle ;
%Shape: Ellipse [id:dp7629833301444369] 
\draw  [fill={rgb, 255:red, 0; green, 0; blue, 0 }  ,fill opacity=1 ] (230.05,49.92) .. controls (230.05,47.27) and (232.37,45.12) .. (235.23,45.12) .. controls (238.09,45.12) and (240.41,47.27) .. (240.41,49.92) .. controls (240.41,52.58) and (238.09,54.72) .. (235.23,54.72) .. controls (232.37,54.72) and (230.05,52.58) .. (230.05,49.92) -- cycle ;
%Shape: Ellipse [id:dp4606570481559854] 
\draw  [fill={rgb, 255:red, 0; green, 0; blue, 0 }  ,fill opacity=1 ] (253.35,77.92) .. controls (253.35,75.27) and (255.67,73.12) .. (258.53,73.12) .. controls (261.39,73.12) and (263.71,75.27) .. (263.71,77.92) .. controls (263.71,80.58) and (261.39,82.72) .. (258.53,82.72) .. controls (255.67,82.72) and (253.35,80.58) .. (253.35,77.92) -- cycle ;
%Straight Lines [id:da5094785320227603] 
\draw    (104.23,51.7) -- (128.4,72.76) ;
%Straight Lines [id:da05770916008182925] 
\draw    (77.47,72.76) -- (87.83,104.35) ;
%Straight Lines [id:da3445024406775993] 
\draw    (77.47,72.76) -- (120.63,103.54) ;
%Straight Lines [id:da4260708802382207] 
\draw    (104.23,51.7) -- (120.63,103.54) ;
%Straight Lines [id:da3749983992664737] 
\draw    (207.61,76.32) -- (234.37,105.92) ;
%Straight Lines [id:da20490392407689806] 
\draw    (235.23,49.92) -- (256.81,75.52) ;
%Straight Lines [id:da9620649003318891] 
\draw    (235.23,49.92) -- (234.37,105.92) ;
%Straight Lines [id:da4147384774986812] 
\draw    (234.37,105.92) -- (258.53,77.92) ;
%Straight Lines [id:da6333682561771077] 
\draw    (87.83,104.35) -- (120.63,103.54) ;
%Shape: Ellipse [id:dp6907215161780947] 
\draw  [fill={rgb, 255:red, 0; green, 0; blue, 0 }  ,fill opacity=1 ] (64.7,313.76) .. controls (64.7,311.36) and (66.82,309.41) .. (69.43,309.41) .. controls (72.04,309.41) and (74.15,311.36) .. (74.15,313.76) .. controls (74.15,316.17) and (72.04,318.12) .. (69.43,318.12) .. controls (66.82,318.12) and (64.7,316.17) .. (64.7,313.76) -- cycle ;
%Shape: Ellipse [id:dp06990409796559027] 
\draw  [fill={rgb, 255:red, 0; green, 0; blue, 0 }  ,fill opacity=1 ] (40.29,332.63) .. controls (40.29,330.23) and (42.4,328.28) .. (45.01,328.28) .. controls (47.62,328.28) and (49.74,330.23) .. (49.74,332.63) .. controls (49.74,335.04) and (47.62,336.98) .. (45.01,336.98) .. controls (42.4,336.98) and (40.29,335.04) .. (40.29,332.63) -- cycle ;
%Shape: Ellipse [id:dp7386885587295666] 
\draw  [fill={rgb, 255:red, 0; green, 0; blue, 0 }  ,fill opacity=1 ] (49.74,360.93) .. controls (49.74,358.53) and (51.85,356.58) .. (54.46,356.58) .. controls (57.07,356.58) and (59.19,358.53) .. (59.19,360.93) .. controls (59.19,363.34) and (57.07,365.29) .. (54.46,365.29) .. controls (51.85,365.29) and (49.74,363.34) .. (49.74,360.93) -- cycle ;
%Shape: Ellipse [id:dp6225104274368898] 
\draw  [fill={rgb, 255:red, 0; green, 0; blue, 0 }  ,fill opacity=1 ] (86.76,332.63) .. controls (86.76,330.23) and (88.87,328.28) .. (91.48,328.28) .. controls (94.09,328.28) and (96.21,330.23) .. (96.21,332.63) .. controls (96.21,335.04) and (94.09,336.98) .. (91.48,336.98) .. controls (88.87,336.98) and (86.76,335.04) .. (86.76,332.63) -- cycle ;
%Shape: Ellipse [id:dp6663581284276052] 
\draw  [fill={rgb, 255:red, 0; green, 0; blue, 0 }  ,fill opacity=1 ] (79.67,360.21) .. controls (79.67,357.8) and (81.78,355.85) .. (84.39,355.85) .. controls (87,355.85) and (89.12,357.8) .. (89.12,360.21) .. controls (89.12,362.61) and (87,364.56) .. (84.39,364.56) .. controls (81.78,364.56) and (79.67,362.61) .. (79.67,360.21) -- cycle ;
%Straight Lines [id:da4545129120419107] 
\draw    (69.43,313.76) -- (91.48,332.63) ;
%Straight Lines [id:da2359616899591136] 
\draw    (45.01,332.63) -- (84.39,360.21) ;
%Straight Lines [id:da4673277327215284] 
\draw    (69.43,313.76) -- (84.39,360.21) ;
%Straight Lines [id:da355547011348066] 
\draw    (54.46,360.93) -- (84.39,360.93) ;
%Shape: Ellipse [id:dp32817445481616003] 
\draw  [fill={rgb, 255:red, 0; green, 0; blue, 0 }  ,fill opacity=1 ] (169.33,314.99) .. controls (169.33,312.53) and (171.46,310.54) .. (174.09,310.54) .. controls (176.72,310.54) and (178.85,312.53) .. (178.85,314.99) .. controls (178.85,317.46) and (176.72,319.45) .. (174.09,319.45) .. controls (171.46,319.45) and (169.33,317.46) .. (169.33,314.99) -- cycle ;
%Shape: Ellipse [id:dp032760002975128266] 
\draw  [fill={rgb, 255:red, 0; green, 0; blue, 0 }  ,fill opacity=1 ] (144.76,334.32) .. controls (144.76,331.85) and (146.88,329.86) .. (149.51,329.86) .. controls (152.14,329.86) and (154.27,331.85) .. (154.27,334.32) .. controls (154.27,336.78) and (152.14,338.77) .. (149.51,338.77) .. controls (146.88,338.77) and (144.76,336.78) .. (144.76,334.32) -- cycle ;
%Shape: Ellipse [id:dp7938909097041279] 
\draw  [fill={rgb, 255:red, 0; green, 0; blue, 0 }  ,fill opacity=1 ] (154.27,363.3) .. controls (154.27,360.83) and (156.4,358.84) .. (159.03,358.84) .. controls (161.65,358.84) and (163.78,360.83) .. (163.78,363.3) .. controls (163.78,365.76) and (161.65,367.76) .. (159.03,367.76) .. controls (156.4,367.76) and (154.27,365.76) .. (154.27,363.3) -- cycle ;
%Shape: Ellipse [id:dp8215363845933056] 
\draw  [fill={rgb, 255:red, 0; green, 0; blue, 0 }  ,fill opacity=1 ] (191.53,334.32) .. controls (191.53,331.85) and (193.66,329.86) .. (196.29,329.86) .. controls (198.91,329.86) and (201.04,331.85) .. (201.04,334.32) .. controls (201.04,336.78) and (198.91,338.77) .. (196.29,338.77) .. controls (193.66,338.77) and (191.53,336.78) .. (191.53,334.32) -- cycle ;
%Shape: Ellipse [id:dp27908257724370267] 
\draw  [fill={rgb, 255:red, 0; green, 0; blue, 0 }  ,fill opacity=1 ] (184.39,362.55) .. controls (184.39,360.09) and (186.52,358.1) .. (189.15,358.1) .. controls (191.78,358.1) and (193.91,360.09) .. (193.91,362.55) .. controls (193.91,365.02) and (191.78,367.01) .. (189.15,367.01) .. controls (186.52,367.01) and (184.39,365.02) .. (184.39,362.55) -- cycle ;
%Straight Lines [id:da7155487556103844] 
\draw    (174.09,314.99) -- (149.51,334.32) ;
%Straight Lines [id:da7875357097733192] 
\draw    (196.29,334.32) -- (159.03,363.3) ;
%Straight Lines [id:da6866828380137033] 
\draw    (174.09,314.99) -- (189.15,362.55) ;
%Straight Lines [id:da7663845629726049] 
\draw    (159.03,363.3) -- (189.15,363.3) ;
%Straight Lines [id:da7493305439803382] 
\draw  [dash pattern={on 4.5pt off 4.5pt}]  (28,269) -- (309.51,269) ;
%Rounded Rect [id:dp6288041697346033] 
\draw   (128.51,313.82) .. controls (128.51,304.88) and (135.77,297.62) .. (144.71,297.62) -- (200.31,297.62) .. controls (209.26,297.62) and (216.51,304.88) .. (216.51,313.82) -- (216.51,362.42) .. controls (216.51,371.37) and (209.26,378.62) .. (200.31,378.62) -- (144.71,378.62) .. controls (135.77,378.62) and (128.51,371.37) .. (128.51,362.42) -- cycle ;
%Rounded Rect [id:dp2683249123871442] 
\draw  [dash pattern={on 0.84pt off 2.51pt}] (192.51,53.62) .. controls (192.51,44.79) and (199.68,37.62) .. (208.51,37.62) -- (258.51,37.62) .. controls (267.35,37.62) and (274.51,44.79) .. (274.51,53.62) -- (274.51,101.62) .. controls (274.51,110.46) and (267.35,117.62) .. (258.51,117.62) -- (208.51,117.62) .. controls (199.68,117.62) and (192.51,110.46) .. (192.51,101.62) -- cycle ;
%Shape: Ellipse [id:dp14430782465816316] 
\draw  [fill={rgb, 255:red, 0; green, 0; blue, 0 }  ,fill opacity=1 ] (164.63,160.88) .. controls (164.63,158.13) and (167.04,155.9) .. (170,155.9) .. controls (172.97,155.9) and (175.37,158.13) .. (175.37,160.88) .. controls (175.37,163.63) and (172.97,165.86) .. (170,165.86) .. controls (167.04,165.86) and (164.63,163.63) .. (164.63,160.88) -- cycle ;
%Shape: Ellipse [id:dp966000521279549] 
\draw  [fill={rgb, 255:red, 0; green, 0; blue, 0 }  ,fill opacity=1 ] (136.9,182.46) .. controls (136.9,179.71) and (139.3,177.48) .. (142.26,177.48) .. controls (145.23,177.48) and (147.63,179.71) .. (147.63,182.46) .. controls (147.63,185.21) and (145.23,187.44) .. (142.26,187.44) .. controls (139.3,187.44) and (136.9,185.21) .. (136.9,182.46) -- cycle ;
%Shape: Ellipse [id:dp459535502922072] 
\draw  [fill={rgb, 255:red, 0; green, 0; blue, 0 }  ,fill opacity=1 ] (147.63,214.83) .. controls (147.63,212.08) and (150.04,209.85) .. (153,209.85) .. controls (155.97,209.85) and (158.37,212.08) .. (158.37,214.83) .. controls (158.37,217.58) and (155.97,219.81) .. (153,219.81) .. controls (150.04,219.81) and (147.63,217.58) .. (147.63,214.83) -- cycle ;
%Shape: Ellipse [id:dp0679967986605976] 
\draw  [fill={rgb, 255:red, 0; green, 0; blue, 0 }  ,fill opacity=1 ] (189.69,182.46) .. controls (189.69,179.71) and (192.09,177.48) .. (195.05,177.48) .. controls (198.02,177.48) and (200.42,179.71) .. (200.42,182.46) .. controls (200.42,185.21) and (198.02,187.44) .. (195.05,187.44) .. controls (192.09,187.44) and (189.69,185.21) .. (189.69,182.46) -- cycle ;
%Shape: Ellipse [id:dp4083117406678365] 
\draw  [fill={rgb, 255:red, 0; green, 0; blue, 0 }  ,fill opacity=1 ] (181.63,214) .. controls (181.63,211.25) and (184.04,209.02) .. (187,209.02) .. controls (189.97,209.02) and (192.37,211.25) .. (192.37,214) .. controls (192.37,216.75) and (189.97,218.98) .. (187,218.98) .. controls (184.04,218.98) and (181.63,216.75) .. (181.63,214) -- cycle ;
%Straight Lines [id:da8530951040501045] 
\draw    (170,160.88) -- (195.05,182.46) ;
%Straight Lines [id:da8809046650564468] 
\draw    (142.26,182.46) -- (194.62,182.14) ;
%Straight Lines [id:da5864641789915306] 
\draw    (170,160.88) -- (187,214) ;
%Straight Lines [id:da7809569688478476] 
\draw    (153,214.83) -- (187,214) ;
%Rounded Rect [id:dp22070959188804629] 
\draw  [dash pattern={on 0.84pt off 2.51pt}] (127.51,162.22) .. controls (127.51,153.05) and (134.95,145.62) .. (144.11,145.62) -- (195.91,145.62) .. controls (205.08,145.62) and (212.51,153.05) .. (212.51,162.22) -- (212.51,212.02) .. controls (212.51,221.19) and (205.08,228.62) .. (195.91,228.62) -- (144.11,228.62) .. controls (134.95,228.62) and (127.51,221.19) .. (127.51,212.02) -- cycle ;
%Straight Lines [id:da3916794236858424] 
\draw    (104.51,118.62) -- (132.51,150.62) ;
%Straight Lines [id:da351424473777178] 
\draw    (143.51,79.62) -- (191.51,80.62) ;
%Shape: Ellipse [id:dp6867371580259339] 
\draw  [fill={rgb, 255:red, 0; green, 0; blue, 0 }  ,fill opacity=1 ] (271.33,315.77) .. controls (271.33,313.31) and (273.46,311.31) .. (276.09,311.31) .. controls (278.72,311.31) and (280.85,313.31) .. (280.85,315.77) .. controls (280.85,318.23) and (278.72,320.23) .. (276.09,320.23) .. controls (273.46,320.23) and (271.33,318.23) .. (271.33,315.77) -- cycle ;
%Shape: Ellipse [id:dp7718241325826568] 
\draw  [fill={rgb, 255:red, 0; green, 0; blue, 0 }  ,fill opacity=1 ] (246.76,335.09) .. controls (246.76,332.63) and (248.88,330.63) .. (251.51,330.63) .. controls (254.14,330.63) and (256.27,332.63) .. (256.27,335.09) .. controls (256.27,337.55) and (254.14,339.55) .. (251.51,339.55) .. controls (248.88,339.55) and (246.76,337.55) .. (246.76,335.09) -- cycle ;
%Shape: Ellipse [id:dp44204940952522953] 
\draw  [fill={rgb, 255:red, 0; green, 0; blue, 0 }  ,fill opacity=1 ] (256.27,364.07) .. controls (256.27,361.61) and (258.4,359.61) .. (261.03,359.61) .. controls (263.65,359.61) and (265.78,361.61) .. (265.78,364.07) .. controls (265.78,366.53) and (263.65,368.53) .. (261.03,368.53) .. controls (258.4,368.53) and (256.27,366.53) .. (256.27,364.07) -- cycle ;
%Shape: Ellipse [id:dp5746445153586432] 
\draw  [fill={rgb, 255:red, 0; green, 0; blue, 0 }  ,fill opacity=1 ] (293.53,335.09) .. controls (293.53,332.63) and (295.66,330.63) .. (298.29,330.63) .. controls (300.91,330.63) and (303.04,332.63) .. (303.04,335.09) .. controls (303.04,337.55) and (300.91,339.55) .. (298.29,339.55) .. controls (295.66,339.55) and (293.53,337.55) .. (293.53,335.09) -- cycle ;
%Shape: Ellipse [id:dp6768762947466718] 
\draw  [fill={rgb, 255:red, 0; green, 0; blue, 0 }  ,fill opacity=1 ] (286.39,363.33) .. controls (286.39,360.87) and (288.52,358.87) .. (291.15,358.87) .. controls (293.78,358.87) and (295.91,360.87) .. (295.91,363.33) .. controls (295.91,365.79) and (293.78,367.79) .. (291.15,367.79) .. controls (288.52,367.79) and (286.39,365.79) .. (286.39,363.33) -- cycle ;
%Straight Lines [id:da6510482745057955] 
\draw    (251.51,334.62) -- (261.03,364.07) ;
%Straight Lines [id:da9481531140969683] 
\draw    (276.09,315.77) -- (291.15,363.33) ;
%Straight Lines [id:da49901988966858946] 
\draw    (261.03,364.07) -- (291.15,363.33) ;
%Rounded Rect [id:dp754818771883102] 
\draw   (230.51,314.6) .. controls (230.51,305.65) and (237.77,298.4) .. (246.71,298.4) -- (302.31,298.4) .. controls (311.26,298.4) and (318.51,305.65) .. (318.51,314.6) -- (318.51,363.2) .. controls (318.51,372.14) and (311.26,379.4) .. (302.31,379.4) -- (246.71,379.4) .. controls (237.77,379.4) and (230.51,372.14) .. (230.51,363.2) -- cycle ;
    
    \end{tikzpicture}
    \caption{A population of networks (bottom) can be seen as a special case of a multilevel network (top) in which each subset of nodes contains the same number of nodes and there are no edges between each subset of nodes.}
    \label{fig:hier_v_pop}
\end{figure}
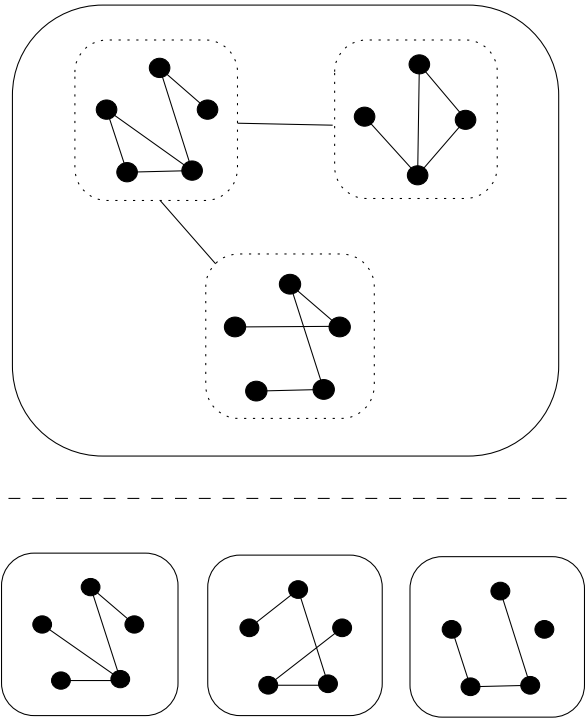

Multilevel networks are networks with a nested hierarchical structure such that nodes may be grouped into subsets of nodes which may further be grouped into subset of subsets of nodes, and so on. It is worth emphasising that the hierarchical nature of a multilevel network corresponds to the grouping of nodes, as opposed to model parameters as might be typical in a Bayesian hierarchical model. A population of networks represents a two-level network such that each subset of nodes correspond to a separate network, with no connections between distinct subsets (see Figure~\ref{fig:hier_v_pop}). \cite{wang2013} proposed ERGMs for multilevel networks, introducing a range of model specifications to account for a range of multilevel structures for two-level networks. \cite{slaughter2016} developed Bayesian hierarchical models for a single group of networks, incorporating covariate information on each network. The framework described in this article can be seen as an extension of this class of models to allow for several groups of networks, allowing for information to be borrowed both within groups and across groups. \cite{yin2020} proposed a mixture of ERGMs to model populations of networks in which the group membership is unknown. \cite{Schweinberger2015} introduced exponential random graph models with local dependence, providing a general framework encompassing multilevel networks (and thus populations of networks) and establishing a central limit theorem for this class of models. 

\subsection*{ERGMs for brain networks}

Exponential random graph models have been applied to resting-state fMRI brain networks (see \cite{Simpson2011} for an early example). \cite{Simpson2013} constructed group-representative networks by taking the mean of the parameter estimates from ERGMs fit to each individual network. \cite{Sinke2016} constructed group-representative networks directly from individual diffusion tensor imaging (DTI) brain networks and then fit Bayesian ERGMs to the resulting group networks. \cite{Obando2017} applied ERGMs to functional connectivity brain networks derived from electroencephalographic (EEG) signals. In each of these approaches, the networks are fit independently from each other and, unlike our approach, there is no borrowing of information across networks.

\subsection*{Other models for populations of networks}

Other statistical network models have recently been extended to handle populations of networks. \cite{Sweet2013} proposed a general framework of hierarchical network models (HNMs), which encompasses the model described in this article. They focus on a hierarchical representation of latent space models \citep{Hoff2002} applied to social networks. \cite{Sweet2014} studied stochastic blockmodel in the HNM framework to infer clusters of nodes shared across networks. \cite{Durante2017} develop an alternative extension of the latent space model \citep{Hoff2002} to populations of networks based on a low-dimensional mixture model representation. \cite{Durante2018} applied this model in the context of groups of networks to test for differences. \cite{mukherjee2017} used graphons to detect clusters among multiple networks within a population (as opposed to clusters within networks). \cite{signorelli2020} use a model-based clustering method based on generalized linear (mixed) models to cluster networks that share certain network properties of interest. 

\section{Model formulation}

\subsection{Exponential random graph models}

The family of exponential random graph models define probability distributions over the space of networks in terms of sets of summary (or sufficient) statistics. We will focus on the case of undirected, binary networks, with $Y_{ij} = Y_{ji} \in \lbrace 0,1 \rbrace$. Let $\mathcal{Y}$ be the range of $\pmb{Y}$, i.e. the set of all possible outcomes. Let $s(\pmb{y}, \pmb{x})$ denote a vector of $p$ summary statistics, such that each component is a function $s_i: \mathcal{Y} \times \mathcal{X}^N \mapsto \mathbb{R}$.

An ERGM is specified by a particular set of $p$ summary statistics and a map $\eta: \Theta \mapsto \mathbb{R}^p $. The probability mass function of $\pmb{Y}$ under the corresponding ERGM is given by
\begin{equation} \label{eq:ergm}
\pi(\pmb{y}|, \pmb{x}, \theta)  = \dfrac{\exp\left\lbrace\eta(\theta)^Ts(\pmb{y}, \pmb{x})\right\rbrace}{Z(\theta)}.
\end{equation}
Here, $\theta \in \Theta \subseteq \mathbb{R}^p$ is a vector of $p$ model parameters that must be estimated from the data and $Z(\theta) = \sum_{\pmb{y}' \in \mathcal{Y}} \exp\left\lbrace\eta(\theta)^Ts(\pmb{y}', \pmb{x})\right\rbrace$ is the normalising constant ensuring the probability mass function sums to one. Given data, that is, a realisation $\pmb{y}$, the goal is to infer which values of $\theta$ best correspond to the data under this distribution. To reduce the notational burden, we will henceforth omit the dependence on the nodal covariates $\pmb{x}$, considering this to be implicit in the specification of the probability distribution.

The Bayesian formulation of ERGMs \citep{Koskinen2004, Caimo2011} augments the definition in (\ref{eq:ergm}) with a prior distribution $\pi(\theta)$ for the model parameters. Given an observation $\pmb{y}$ of the network, inference is performed by analysing the posterior distribution $\pi(\theta|\pmb{y})$:
\begin{equation}
\begin{split}
\pi(\theta|\pmb{y}) &= \dfrac{\pi(\pmb{y}|\theta)\pi(\theta)}{\pi(\pmb{y})} \\
&= \dfrac{\exp\left\lbrace\eta(\theta)^Ts(\pmb{y}, \pmb{x})\right\rbrace\pi(\theta)}{Z(\theta)\pi(\pmb{y})}, 
\end{split}
\end{equation}
where $\pi(\pmb{y}) = \int_\Theta \pi(\pmb{y}|\theta)\pi(\theta) d\theta$ is the model evidence. The posterior distribution is generally not available in a closed-form expression. This is due to two properties of the posterior: first, the (standard) intractability of the model evidence, and second, the intractability of the likelihood via the intractable normalising constant $Z(\theta)$. Posterior distributions with these two sources of intractability are referred to as \textit{doubly-intractable}. We explore how to sample from this posterior in Section~\ref{inference}.

\subsection{A framework for populations of networks} \label{multilevel}

The Bayesian exponential random graph model described above provides a flexible family of distributions for a \textit{single} network. Our aim is to extend this to a model for a \textit{population} of networks by representing each network as a separate ERGM within a Bayesian multilevel (or hierarchical) model. By pooling information across individual networks, this approach allows us to characterise the distribution of the whole population. Let $\bm{Y} = (\pmb{Y}^{(1)}, \dots, \pmb{Y}^{(n)})$ be a set of $n$ networks. Identify each network $\pmb{Y}^{(i)}$ with its own \textit{individual-level} (and vector-valued) ERGM parameter $\theta^{(i)}$. Write $\bm{\theta} = (\theta^{(1)}, \dots, \theta^{(n)})$ for the set of individual-level parameters. 

\begin{figure}
	\centering
	\large{\begin{tikzpicture}[%
		->,
		>=stealth,
		node distance=1cm,
		minimum size = 1.4cm,
		pil/.style={
			->,
			thick,
			shorten =2pt,
			draw = black}
		]
		\node[circle, draw, double, fill = red] (Y11) {$\pmb{Y}^{(1)}$};
		\node[right=1cm of Y11] (dots1) {$...$};
		\node[circle, draw, double, right=1cm of dots1, fill = red] (Y1n) {$\pmb{Y}^{(n)}$};
		\node[circle, draw, above=of Y11, fill = red,  fill opacity = 0.5, text opacity=1] (theta11) {$\theta^{(1)}$};
		\node[circle, draw, above=of Y1n, fill = red,  fill opacity = 0.5, text opacity=1] (theta1n) {$\theta^{(n)}$};
		\node[right=1cm of theta11] (dots3) {$...$};
		
		\draw [->] (theta11) -- (Y11);
		\draw [->] (theta1n) -- (Y1n);
		
		\node[circle, draw, above=of dots3, fill = red,  fill opacity = 0.5, text opacity=1] (mu1) {$\phi$};
		\draw [->] (mu1) -- (theta11);
		\draw [->] (mu1) -- (theta1n);
	\end{tikzpicture}}
	\caption{A diagrammatic representation of the hierarchical framework. Each network $\pmb{Y}^{(i)}$ is modelled as an exponential random graph with individual-level parameter $\theta^{(i)}$. In turn, each $\theta_i$ is assumed to come from a common population-level distribution with hyperparameter $\phi$.}
    \label{fig:groupHier}
\end{figure}
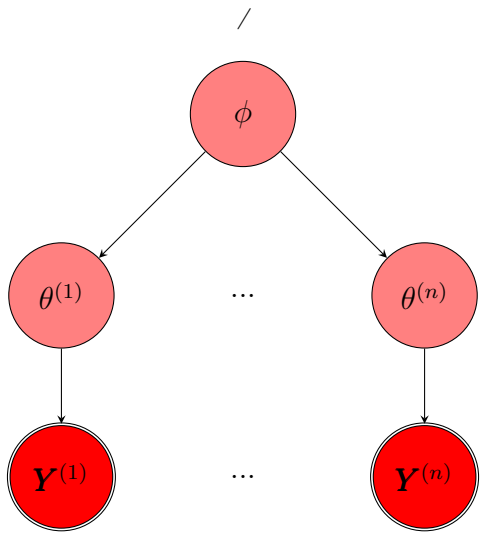

We model each individual network $\pmb{Y}^{(i)}$ as an exponential random graph with model parameter $\theta^{(i)}$. Importantly, each individual ERGM must consist of the same set of $p$ summary statistics $s(\cdot)$. We assume that, conditional on their respective individual-level parameters, the $\pmb{Y}^{(i)}$ are independent. We place a population-level prior distribution $\pi(\bm{\theta}|\phi)$ on the individual-level ERGM parameters. We again assume conditional independence of the $\theta^{(i)}$ conditional on the hyperparameter $\phi$, and complete the model with a hyperprior $\pi(\phi)$ on $\phi$. See Figure \ref{fig:groupHier} for a diagrammatic representation of the full model.

\subsubsection*{Prior specification}

The choice of prior distributions for single network Bayesian ERGMs has yet to be studied in any great detail. The appropriate setting of priors is a challenging task due to the typically high levels of dependence between parameters \cite{Koskinen2013}. Studies thus far have generally assumed (flat) multivariate normal prior distributions on the model parameters \citep{Caimo2011, Sinke2016, Thiemichen2016}. Through the choice of the prior distribution $\pi(\phi)$, we can encode additional information on the population of networks, such as group structure. For conceptual and computational simplicity, we will also assume multivariate normal priors, though other prior specifications warrant further investigation.

Let $g_i \in \left\lbrace 1, \dots, J \right\rbrace$ denote the group membership of network $i$. On each individual-level parameters, we place a multivariate normal prior with group-level hyperparameters $(\mu^{(g_i)}, \pmb{\Sigma}^{(g_i)}_\theta)$: 
\begin{equation}
\theta^{(i)} \sim \mathcal{N}(\mu^{(g_i)}, \pmb{\Sigma}^{(g_i)}_\theta), ~~~i = 1, \dots, n.
\end{equation}

We complete the model by specifying (partially) conjugate hyperpriors on the group-level parameters: 
\begin{equation}
\begin{split}
\mu^{(j)} &\sim \mathcal{N}(\mu^{pop}, \pmb{\Sigma}_\mu) \\
\pmb{\Sigma}^{(j)}_\theta &\sim \mathcal{W}^{-1}(\nu_\theta, \pmb{\Psi}_\theta) \\
\mu^{pop} &\sim \mathcal{N}(\mu_0, \pmb{\Lambda}) \\
\pmb{\Sigma}_\mu &\sim \mathcal{W}^{-1}(\nu_\mu,\pmb{\Psi}_\mu) 
\end{split}
\end{equation}
This allows information to be pooled across the $J$ groups via the population-level hyperparameter $\mu^{pop}$ while also allowing for differences in the group-level means.

\section{Posterior computation} \label{inference}

The double-intractability of the ERGM posterior distribution means that standard MCMC schemes such as the Metropolis algorithm are not suitable. This is due to the presence of the intractable normalising constants $Z(\theta^{(i)}$ in the denominator, rendering calculation of the Metropolis acceptance rates computationally infeasible. Several methods have been proposed in recent years to perform Bayesian inference in the presence of intractable normalising constants (see \cite{Park2018} for a review). We focus here on the exchange algorithm \citep{Murray2006}, which was employed to generate posterior samples for single-network Bayesian ERGMs by \cite{Caimo2011}. We first recap the exchange algorithm in the context of Bayesian ERGMs before describing a \textit{exchange-within-Gibbs} scheme to generate samples from the joint posterior.

Consider a Metropolis update for a single-network Bayesian ERGM. The acceptance probability for a proposal $\theta'$ from current value $\theta$ requires evaluation of the ratio $Z(\theta) / Z(\theta')$, which is computationally intractable. The exchange algorithm is an MCMC scheme designed to circumvent this obstacle. This is achieved by introducing an auxiliary variable $\pmb{y}' \sim \pi(\cdot|\theta')$, i.e. a network drawn from the same exponential random graph model with parameter $\theta'$. 

The algorithm targets an augmented posterior
\begin{equation}
\pi(\theta, \theta', \pmb{y}'|\pmb{y}) \propto \pi(\theta|\pmb{y})h(\theta'|\theta)\pi(\pmb{y}'|\theta') \label{eq:augmentedPosterior}
\end{equation}
where $\pi(\theta|\pmb{y})$ is the original (target) posterior, $h(\theta'|\theta)$ is an arbitrary, normalisable proposal function, and $\pi(\pmb{y}'|\theta)$ is the likelihood of the auxiliary variable. For simplicity, we assume $h(\theta'|\theta)$ to be symmetric. Each of the three terms on the right-hand side of Eq. \eqref{eq:augmentedPosterior} can be normalised, so the left-hand side is well-defined as a probability distribution. 

The algorithm proceeds as follows. At each iteration, first perform a Gibbs' update of $(\theta', \pmb{y}')$ by drawing $\theta' \sim h(\cdot | \theta)$ followed by $\pmb{y}' \sim \pi(\cdot|\theta')$. Next, \textit{exchange} $\theta$ and $\theta'$ with probability $\min(1, AR(\theta', \theta, \pmb{y}, \pmb{y}'))$, where
\begin{equation}
\begin{split}
AR(\theta', \theta, \pmb{y}, \pmb{y}') &= \dfrac{\pi(\theta'|\pmb{y})}{\pi(\theta|\pmb{y})} \cdot \dfrac{\pi(\pmb{y}'|\theta)}{\pi(\pmb{y}'|\theta')} \\
&= \dfrac{\exp\left\lbrace\theta'^Ts(\pmb{y})\right\rbrace\pi(\theta')}{\exp\left\lbrace\theta^Ts(\pmb{y})\right\rbrace\pi(\theta)}\dfrac{Z(\theta)}{Z(\theta')} \cdot \dfrac{\exp\left\lbrace\theta^Ts(\pmb{y}')\right\rbrace}{\exp\left\lbrace\theta'^Ts(\pmb{y}')\right\rbrace}\dfrac{Z(\theta')}{Z(\theta)} \\
&= \exp\left\lbrace[\theta' - \theta]^T[s(\pmb{y}) - s(\pmb{y}')]\right\rbrace\frac{\pi(\theta')}{\pi(\theta)}
\end{split} \label{eq:exAR}
\end{equation}
Crucially, the ratio of intractable normalising constants cancel out, and so this acceptance ratio can indeed be evaluated. The stationary distribution of the Markov chain constructed through this scheme is $\pi(\theta, \theta', \pmb{y}'|\pmb{y})$ \citep{Murray2006}. Thus, by marginalising out $\theta'$ and $\pmb{y}'$, the algorithm yields samples from the desired posterior, namely $\pi(\theta|\pmb{y})$.

\begin{algorithm}
\caption{The exchange algorithm for a Bayesian ERGM \citep{Caimo2011} \label{alg:exchange} }

\begin{algorithmic}
\Require number of MCMC iterations $K$, initial value $\theta_0$
\For{$k = 1, \dots, K$}
\State - draw $\theta' \sim h(\cdot|\theta_{k-1})$ 
\State - draw $\pmb{y}' \sim \pi(\cdot|\theta')$

\State - set $\theta_k = \theta'$ with probability $\min \left(1, AR(\theta', \theta_{k-1}, \pmb{y}, \pmb{y}') \right)$ \Comment{See Eq. (\ref{eq:exAR})}
\State - else, set $\theta_k = \theta_{k-1}$
\EndFor
\end{algorithmic}
\end{algorithm}

At each iteration, the exchange algorithm requires a sample $\pmb{y}'$ from the ERGM $\pi(\cdot|\theta')$ in order to compute the acceptance ratio. Although perfect sampling for ERGMs is possible, it is computationally impractical except for a few special cases \citep{Butts2018}. A pragmatic alternative, employed in \cite{Caimo2011} and \cite{wang2014}, is to use the final iteration of a Metropolis-Hastings algorithm as an approximate sample from $\pi(\cdot|\theta')$ \citep{Hastings1970, Hunter2008a}. A theoretical justification of this approach is given by \cite{Everitt2012}: under certain conditions, despite using an approximate sample, the algorithm nevertheless targets an approximation to the correct posterior distribution. Further, this approximation improves as the number of iterations of the inner MCMC increases.

\subsection{The exchange-within-Gibbs algorithm}

We now extend the exchange algorithm in order to generate samples from our full posterior on a population of networks. As the name suggests, the exchange-within-Gibbs algorithm combines the exchange algorithm with the Gibbs sampler \citep{Geman1984} to produce samples from the desired posterior. Note that we can treat the unknown parameters of the model $(\pmb{\theta}, \phi)$ as components of a single multi-dimensional parameter. We iteratively sample each component from its conditional distribution given the remaining components.

The full exchange-within-Gibbs scheme is outlined in Algorithm \ref{alg:exchangewithinGibbs}. Since each step samples from the respective full conditional distribution, the algorithm ensures that the stationary distribution of the resulting Markov chain is indeed the joint posterior $\pi(\pmb{\theta}, \phi|\bm{y})$ \citep{Tierney1994}. As with the exchange algorithm for the single-network Bayesian ERGM, the most computationally expensive step is sampling $\pmb{y}'$ from $\pi(\cdot|\theta')$, i.e. simulating an exponential random graph with parameter $\theta'$. Moreover, this step must be performed for each of the individual-level parameter $\theta^{(i)}$ updates. Thus, the computational cost of each iteration increases linearly with the number of networks in the data. However, since (conditional on $\phi$) the $\theta^{(i)}$ are independent, these updates may be performed in parallel. Therefore, with access to a sufficient number of computing cores, the actual computational time per iteration typically increases sub-linearly with the number of networks.

\begin{algorithm}
\caption{\textcolor{black}{The exchange-within-Gibbs algorithm for a multilevel Bayesian ERGM} \label{alg:exchangewithinGibbs}} 

\begin{algorithmic}
\Require number of MCMC iterations $K$, initial values $(\phi_0, \theta^{(1)}_0, \dots, \theta^{(n)}_0)$
\For{$k = 1, \dots, K$}
\State - draw $\phi_k \sim \pi(\cdot|\pmb{\theta}_{k-1})$

\For{$i = 1, \dots, n$}
\State - update $\theta^{(i)}_k$ via the exchange algorithm 

\EndFor
\EndFor
\end{algorithmic}

\end{algorithm}

\subsubsection*{Choice of parametrisation: centering vs. non-centering}

The parametrisation of general multilevel models in the context of MCMC computation has been studied in some detail \citep{Gelfand1995, Papaspiliopoulos2003, Papaspiliopoulos2007, Yu2011}. Here, we discuss the two most commonly used parametrisations: the `centered' and the `non-centered'. The parametrisation presented thus far is known as the centered parametrisation (CP) \cite{Gelfand1995, Papaspiliopoulos2007}, in which the group-level parameters ($\mu, \pmb{\Sigma}_\theta$) are independent of the data $\bm{Y}$:
\begin{equation}
\begin{split}
\pmb{Y}^{(i)} &\sim ERGM(\theta^{(i)}), ~~~ i = 1, \dots, n \\
\theta^{(i)} &\sim \mathcal{N}(\mu, \pmb{\Sigma}_\theta), ~~~ i = 1, \dots, n. 
\end{split}
\end{equation}
In contrast, the \textit{non-centred} parametrisation (NCP) can be written as follows:

\begin{equation}
\begin{split}
\pmb{Y}^{(i)} &\sim ERGM(\mu + \tilde{\theta}^{(i)}), ~~~ i = 1, \dots, n \label{eq:NCPlike} \\
\tilde{\theta}^{(i)} &\sim \mathcal{N}(0, \pmb{\Sigma}_\theta), ~~~ i = 1, \dots, n.
\end{split}
\end{equation}
The transformation $\theta^{(i)} = \mu + \tilde{\theta}^{(i)}$ confirms the equivalence of the two parametrisations. Note that the group-level parameter $\mu$ enters the likelihood directly in \eqref{eq:NCPlike}, so the conditional distribution of $\mu$ given the remaining parameters has an intractable normalising constant. As above, this can be dealt with via an exchange update, in this case requiring simulation of $n$ networks for the normalising constants to cancel in the acceptance ratio.

The centred parametrisation and the non-centred parametrisation tend to be complementary: when one performs poorly, the other tends to perform better \citep{Papaspiliopoulos2007}. However, when the parameters of interest are the group-level parameters $(\mu, \pmb{\Sigma}_\theta)$, it is possible to combine both approaches using an ancillarity-sufficiency interweaving strategy (ASIS; \cite{Yu2011}). ASIS works by combining the updating schemes of the CP and NCP approaches. The ASIS algorithm for a multilevel Bayesian ERGM is described in Algorithm \ref{alg:ASIS}.

\begin{algorithm}
    \caption{The ancillarity-sufficiency interweaving strategy (ASIS) algorithm for a multilevel Bayesian ERGM} \label{alg:ASIS}
    \begin{algorithmic}

    \Require number of MCMC iterations $N$, initial values $(\mu_0, \tilde{\theta}^{(1)}_0, \dots, \tilde{\theta}^{(n)}_0)$
    \For{$k = 1, \dots, K$}

        \State Draw $\pmb{\Sigma}_{\theta,k} \sim \pi(\cdot|\mu_{k-1}, \tilde{\pmb{\theta}}_{k-1})$
        \For{$i = 1, \dots, n$}
            \State Update $\theta^{(i)}_{k - 0.5}$ using the exchange algorithm 

        \EndFor \vspace{0.2cm}

        \State Draw centered $\mu_{k-0.5} \sim \pi(\cdot|\pmb{\theta}_{k-0.5}, \pmb{\Sigma}_{\theta,k})$
        \For{$i = 1, \dots, n$}
            \State Set $\tilde{\theta}^{(i)}_k = \theta_{k-0.5}^{(i)} - \mu_{k-0.5}$ 
        \EndFor \vspace{0.2cm}
        \State Update non-centered $\mu_k$ (see \eqref{eq:NCPlike}) via the exchange algorithm 

    \EndFor
\end{algorithmic}
\end{algorithm}

\subsubsection*{Proposal adaptation}

We use multivariate normal random walk proposals in the respective exchange updates of both $\theta^{(i)}$ and $\mu$, for example
\begin{equation}
h(\theta' | \theta_{k-1}) = \mathcal{N}(\theta_{k-1}, \Sigma)
\end{equation}
The choice of the the proposal covariance matrix $\Sigma$ is crucial to the overall efficiency of the MCMC algorithm; we wish to make large proposals that are likely to be accepted in order to explore the posterior in as few iterations as possible. A common approach to tuning covariance proposals for a wide range of random walk based algorithms, including Metropolis-within-Gibbs, is to target an acceptance rate close to 0.234, with acceptance rates between 0.1 and 0.5 often yielding satisfactory results \citep{roberts1997weak, roberts2001optimal, roberts2009examples}. Since manual tuning of the $n + 1$ proposal covariance matrices would be impractical, we instead implement an adaptive proposal scheme.

For each proposal, we use a version of the adaptive Metropolis algorithm \citep{haario2001adaptive} considered by \cite{roberts2009examples}. Specifically, for the first 1000 iterations, we adapt every 20 iterations, with proposals of the form
\begin{equation}
h_k(\theta' | \theta_{k-1})  = (1-\beta) N \left( \theta_{k-1}, (2.38)^{2} \delta_{k} \Sigma_{k} / p \right) + \beta N\left(\theta_{k-1}, (0.1)^{2} \delta_{k} I_{p} / p \right),
\end{equation}
where $\Sigma_k$ is the sample covariance matrix of the posterior samples $(\theta_{1},\dots,\theta_{k-1})$ and $\delta_k$ is an additional scaling factor that is varied to control the magnitude of the proposals. Following \cite{roberts2009examples}, we set $\beta = 0.05$. The role of $\Sigma_k$ is to adapt the direction of the proposals to the MCMC run so far, while $\delta_k$ serves to target an acceptance rate of 0.234. Specifically, we start with $\delta_1 = 1$ and increase (resp. decrease) $\log(\delta_k)$ by $\min (0.5, 1 / \sqrt(k)$ if the acceptance rate was below (resp. above) 0.234 in the previous 20 iterations.

\subsection{Posterior predictive assessment}

Having produced a sufficient number of samples from the posterior distribution, we then assess whether the model adequately describes the data. Since determining the distribution of appropriate test quantities is difficult, assessing such goodness-of-fit for ERGMs is typically performed graphically \citep{Hunter2008b}. For a single ERGM fit, one can simulate a large number of networks from the fitted model and compare these `posterior predictive networks' to the observed network. This comparison is usually done via a set of network metrics. If a model fits the data well then the network metrics of the posterior predictive networks should be similar to those of the observed network. 

For a population of networks, we can apply the same principles. To do so, we choose uniformly at random $S$ values from the posterior samples of the group-level mean parameters. For each value, we simulate a network from $\pi(\cdot|\mu^{(s)})$. We can then compare these posterior predictive networks to the observed networks based on a set of network metrics. For this purpose, we will use three important network metric distributions that are not explicitly modelled, namely degree distribution, geodesic distance distribution (length of shortest paths) and edge-wise shared partners distribution. 

\section{Results}

To illustrate our method, we apply it to a set of simulated networks, demonstrating that it is capable of recovering the ground truth. We also apply our method to resting-state fMRI networks from the Cam-CAN project, a study on healthy ageing \citep{Shafto2014}, finding that differences in network structure between a group of young individuals and and a group of old individuals can largely be explained by a reduction in inter-hemispheric connections in the old. The R scripts used to generate these results can be found at \url{https://github.com/brieuclehmann/multibergm-scripts}.

\subsection{Simulation}

We generated sets of 30-node networks with nodes split into two `hemispheres' of 15 nodes each. We simulated the networks from an exponential random graph model with three terms: total number of edges (`edges'), total number of edges between nodes in the same hemisphere (`nodematch.hemisphere'), and the geometrically-weighted edgewise-shared partner (GWESP) statistic (`gwesp.fixed.0.9'). The GWESP statistic of a network $y$ is a measure of clustering and is given by:
\begin{equation}
    GWESP(\pmb{y}) = e^\tau \sum_{w=1}^N \lbrace 1 - (1 - e^{-\tau})^w \rbrace EP_w(\pmb{y}),
\end{equation}
where $EP_w(y)$ is the number of connected node pairs having exactly $w$ shared partners and $\tau$ is a decay parameter, which we fix at $\tau = 0.9$. The decay parameter attenuates the effect of the number of higher-order edgewise shared partners relative to lower-order edgewise shared partners. 

To illustrate the method, we first simulate networks for a single group with $n = 10, 20, 50$ networks, before moving onto an example with two groups, such that $n = 20, 40, 100$ and $n_1 = n_2$, i.e. both groups have the same number of networks. 

\subsubsection*{Single group}

To simulate the networks, we first generated individual-level parameters $\theta_i \sim \mathcal{N}(\mu, \Sigma), ~ i = 1, \dots, n$ where
\begin{align}
    \mu &= (-3, 0.5, 0.5)^T \\
    \Sigma &= \frac{1}{50}\begin{pmatrix}
1 & -0.5 & 0 \\
-0.5 & 0.5 & 0 \\
0 & 0 & 0.5
\end{pmatrix}.
\end{align}
We then used the \texttt{ergm} R package \citep{Hunter2008a} to simulate $n$ networks $\pmb{y}_i \sim p(\cdot|\theta_i), ~ i = 1, \dots, n$. The simulation procedure is based on an MCMC algorithm, initialised at a network with the prescribed number of nodes and covariates (in this case, hemisphere labels). With these simulated networks, we applied our exchange-within-Gibbs algorithm with ASIS (Algorithm \ref{alg:ASIS}) to generate 12,000 posterior samples, adapting the random-walk proposals for the first 1,000 iterations, and discarding the first 2,000 as burn-in. 

\begin{figure}
    \centering
    \includegraphics[width=0.8\textwidth]{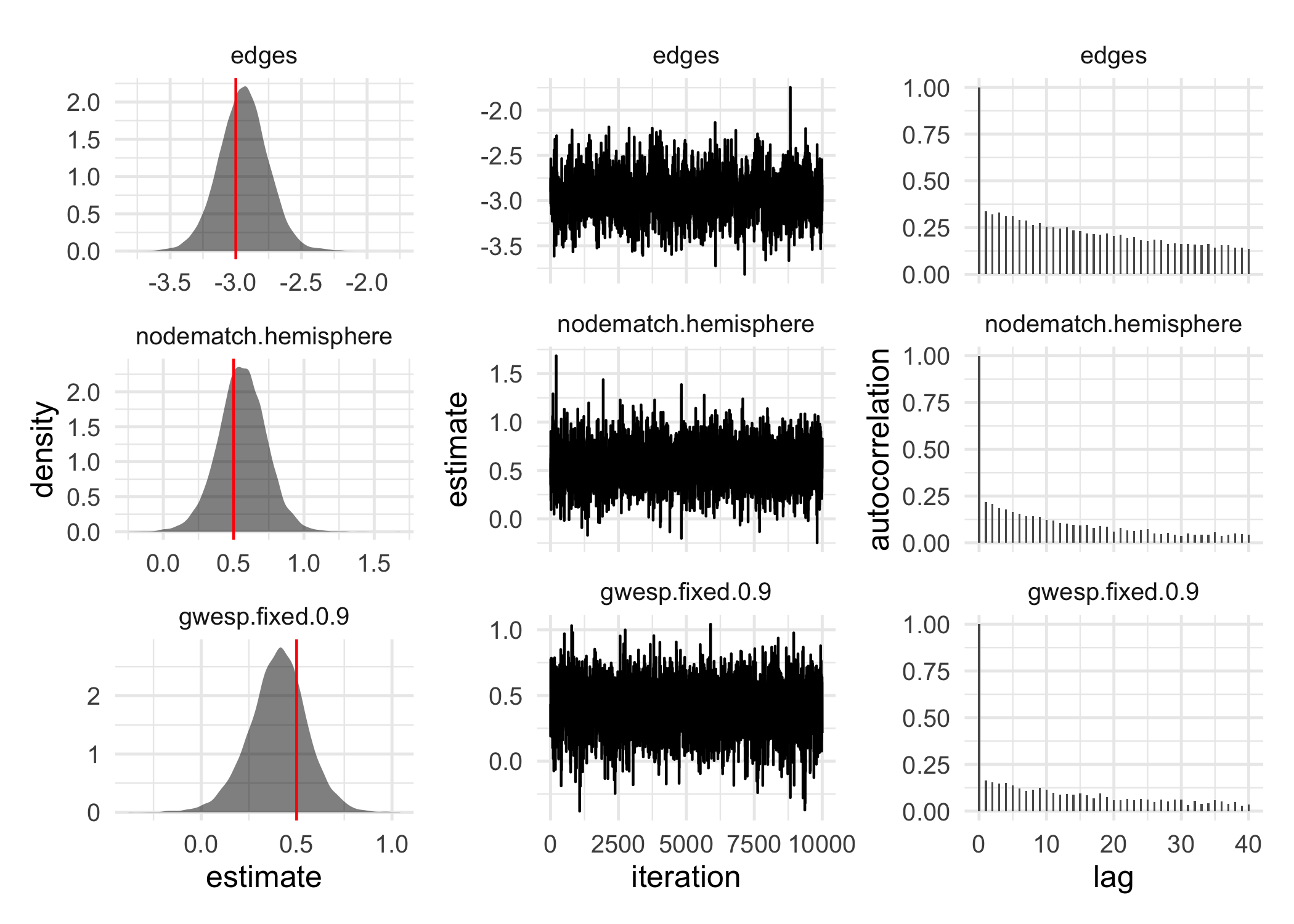}
    \caption{Posterior samples produced by the exchange-within-Gibbs algorithm for the group-level mean parameter, $\mu$, of a single group of ten simulated networks. The true value of $\mu$ is indicated by the red line.}
    \label{fig:single_n10}
\end{figure}

\begin{figure}
    \centering
    \includegraphics[width=0.8\textwidth]{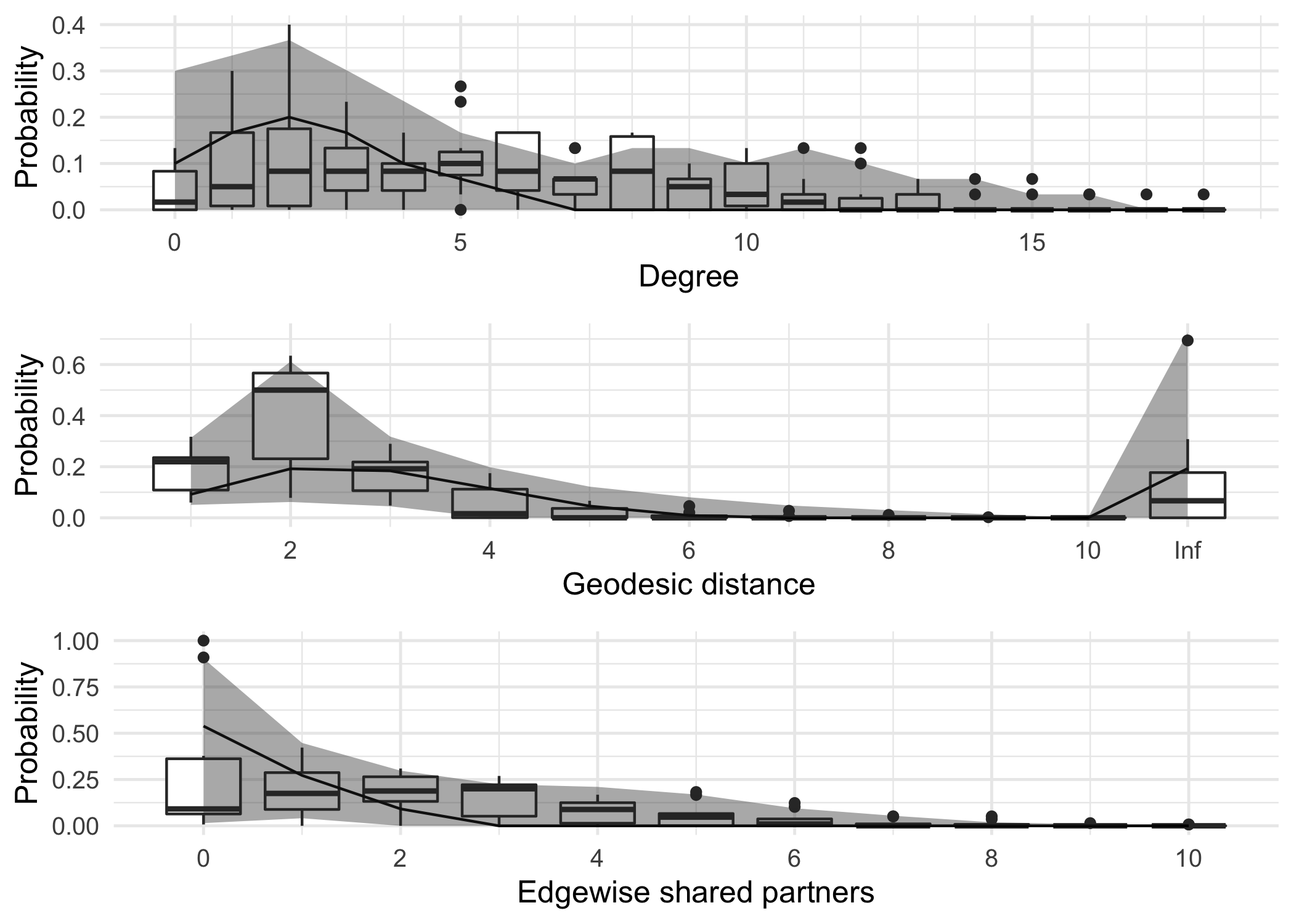}
    \caption{Graphical goodness-of-fit assessment for a single group of ten simulated networks. The box plots correspond to the simulated networks, while the ribbons represent 90\% credible intervals corresponding to the posterior predictive networks. Note that a geodesic distance of infinity between two nodes means that there is no path connecting the nodes.}
    \label{fig:single_n10_gof}
\end{figure}

Figure \ref{fig:single_n10} displays summaries of the posterior samples for the group-level mean parameter $\mu$ of the model fit to $n=10$ networks. The true value of $\mu$ is covered by the posterior density, while the trace and autocorrelation plots indicate that the MCMC has mixed well. To assess the goodness-of-fit, we generated $S=100$ networks from the model at posterior samples of $\mu$ chosen uniformly at random. Figure \ref{fig:single_n10_gof} shows the degree distribution, geodesic distance distribution and edgewise shared partner distribution of these simulated networks against those to which the model was fit. 

\begin{figure}
    \centering
    \includegraphics[width=0.9\textwidth]{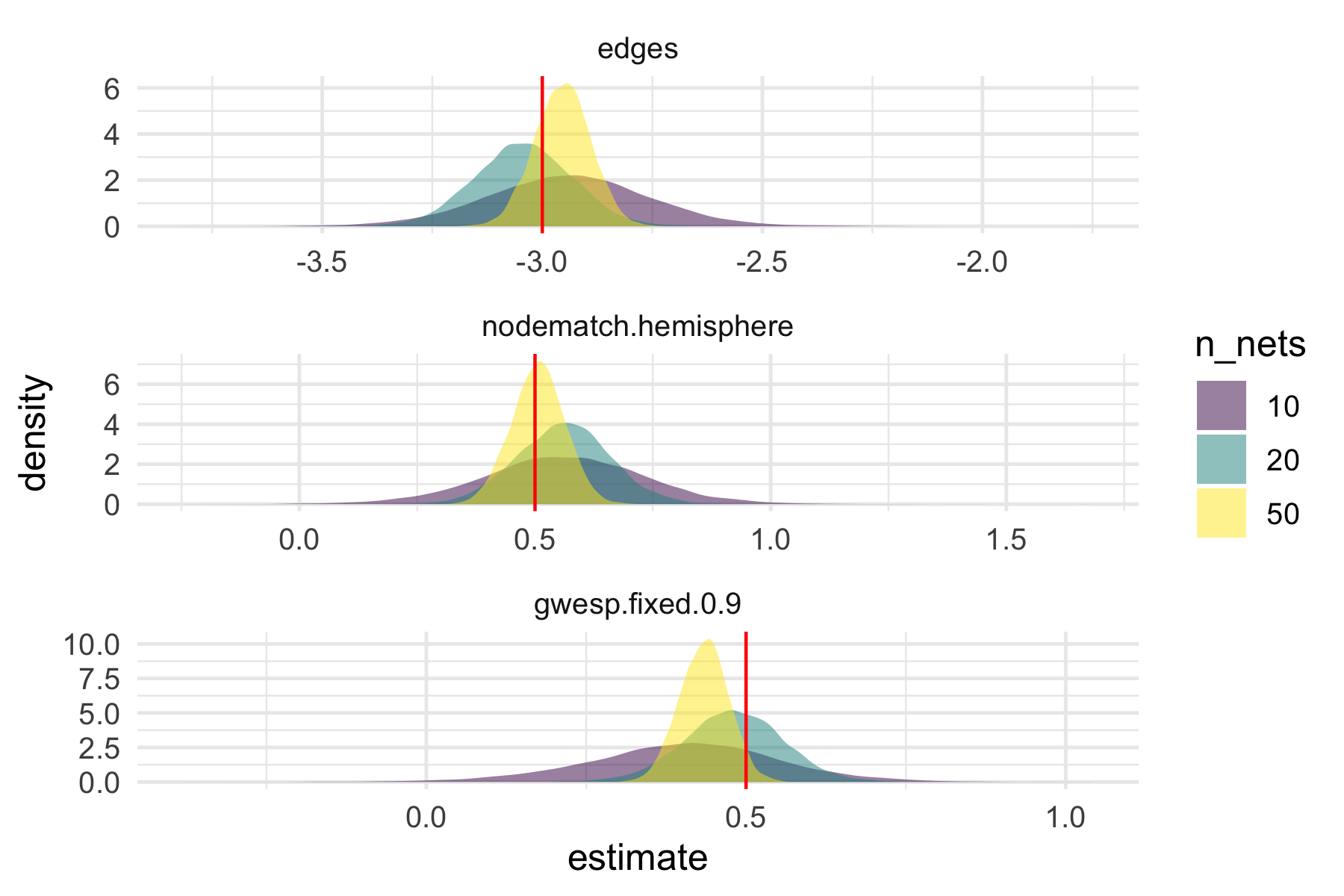}
    \caption{Posterior density plots for the group-level mean parameters with varying number of networks. As the number of networks increases, the posterior concentrates around the true value, depicted by the red vertical line.}
    \label{fig:single_combi}
\end{figure}

To complete our analysis of a single group of networks, we compare the density of the posterior samples between groups of size $n = 10, 20, 50$. Figure~\ref{fig:single_combi} illustrates how the posterior samples of $\mu$ concentrates around the true value as the number of networks in the group increases.

\subsubsection*{Two groups}

We now consider a multilevel setting in which the networks are split into two distinct groups $j = 1, 2$, each with their own group-level mean parameters $\mu^{(j)}$ but with common group-level covariance $\Sigma$. As above, we first generated individual-level parameters $\theta_i \sim \mathcal{N}(\mu^{(g_i)}, \Sigma), ~ i = 1, \dots, n$, where $g_i \in \lbrace 1, 2 \rbrace$ denotes the group membership of the $i^{th}$ network, and then simulated networks $\pmb{y}_i \sim p(\cdot|\theta_i), ~ i = 1, \dots, n$. We considered a range of numbers of networks, $n = 10, 20, 50$ each of the two groups. The true values were
\begin{align}
    \mu^{(1)} &= (-3, 0.5, 0.5)^T \\
    \mu^{(2)} &= (-2.6, 0.5, 0.2)^T \\
    \Sigma &= \frac{1}{50}\begin{pmatrix}
1 & -0.5 & 0 \\
-0.5 & 0.5 & 0 \\
0 & 0 & 0.5
\end{pmatrix}.
\end{align}

\begin{figure}
    \centering
    \includegraphics[width=\textwidth]{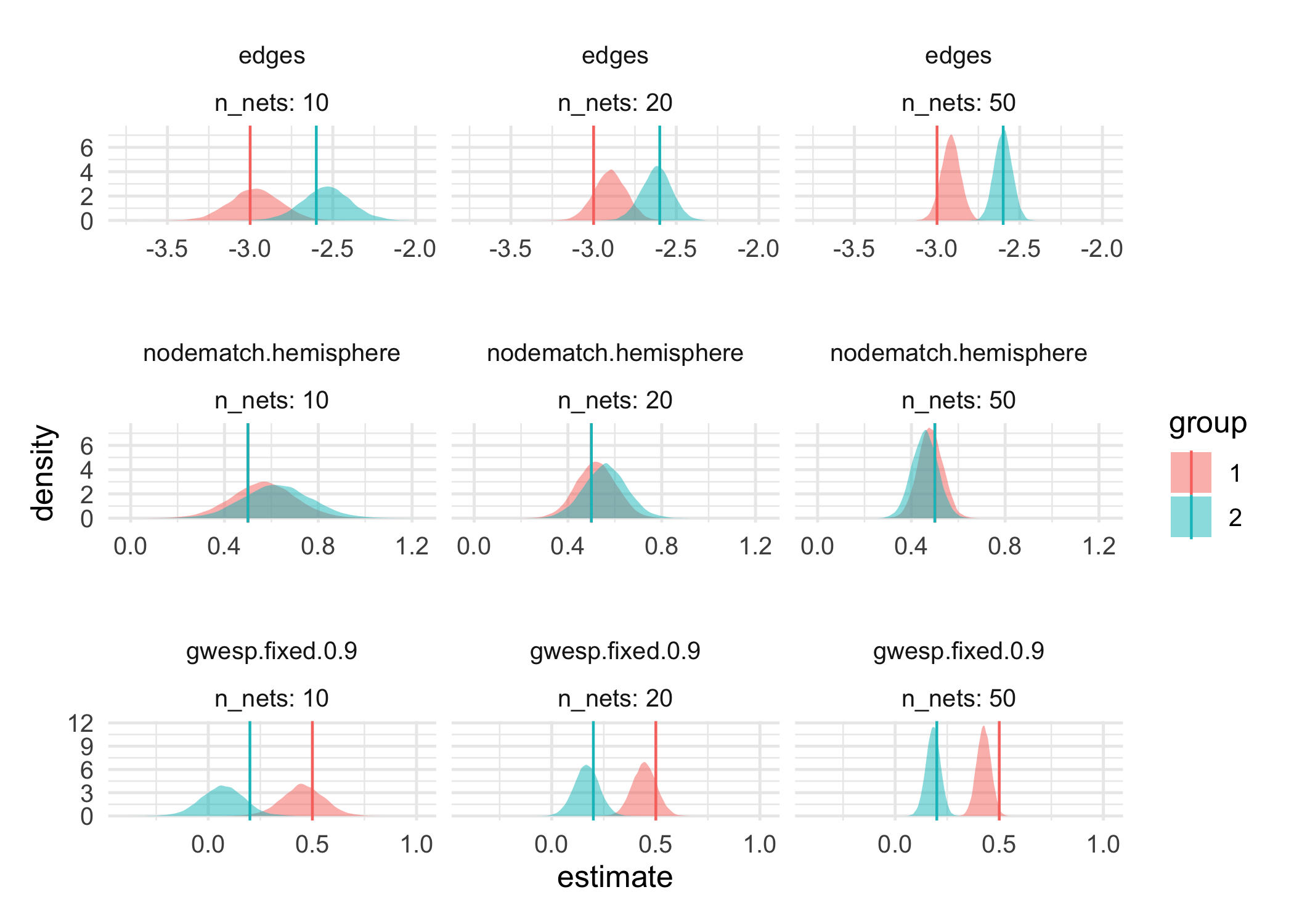}
    \caption{Posterior density plots for a two-group model with $n = 10,20,50$ simulated networks in each group. }
    \label{fig:twogrp_combi}
\end{figure}

Figure~\ref{fig:twogrp_combi} shows the density of the posterior samples for the group-level parameters $(\mu^{(1)}, \mu^{(2)})$ for increasing number of networks $n$ per group. We see that, as in the single-group setting, the posteriors concentrate around the true values for each group as the number of networks increases. 

\subsection{Application to human functional connectivity brain networks}

We now turn our attention to a real data example: networks derived from resting-state fMRI scans of human brains from the Cambridge Centre for Ageing and Neuroscience (Cam-CAN) research project \cite{Shafto2014}, a study on the effect of healthy ageing on cognitive and brain function. The Cam-CAN dataset consists of a range of cognitive tests and functional neuroimaging experiments for approximately 650 healthy individuals aged 18-87. Our aim will be to compare the functional connectivity structure between the 100 youngest individuals, aged 18–33, and the 100 oldest individuals, aged 74–87. 

Full details of data collection and preprocessing can be found in \cite{lehmann2021}. To summarise, both structural (T1 and T2) and eyes-closed, resting-state fMRI scans (261 volumes, lasting 8min 40s) were acquired for each individual. The fMRI scans were motion-corrected and co-registered to the respective structural scans and then mapped to the common Montreal Neurological Institute (MNI) template to ensure comparability across individuals. The fMRI time series were then extracted from 90 cortical and subcortical regions of interest (ROIs) from the AAL atlas \citep{TzourioMazoyer2002} and adjusted for various confounds using the optimised pipeline of \cite{Geerligs2017}.

To construct networks for each individual, we followed a thresholded correlation matrix approach. For individual $i$, we computed the pairwise Pearson correlation between each of the $N = 90$ preprocessed time series, yielding a $N \times N$ correlation matrix $\pmb{C}^{(i)}$. We then applied a threshold $r$ to $\pmb{C}^{(i)}$ to produce an $N \times N$ adjacency matrix, $\pmb{A}^{(i)}$, with entries:
\begin{equation}
\pmb{A}^{(i)}_{kl} = \begin{cases} 
1 &\text{if } \pmb{C}^{(i)}_{kl} \geq r \quad k,l=1,\dots,N\\
0 &\text{otherwise.}
\end{cases}
\end{equation}
The adjacency matrix defines an individual's network, $\pmb{y}^{(i)}$, with an edge between nodes $k$ and $l$ if and only if $\pmb{A}^{(i)}_{kl}=1$. The threshold $r$ was chosen to yield an average node degree of 3 across all the networks, as recommended by \cite{Fallani2017}.

We model the resulting population of $n = 200$ networks using the framework described in Section~\ref{multilevel} with an exponential random graph model with four terms: total number of edges (`edges'), total number of edges between nodes in the same hemisphere (`nodematch.hemisphere'), total number of edges between homotopic nodes (mirror ROIs in each hemisphere; `nodematch.homotopy') and the geometrically-weighted edgewise-shared partner (GWESP) statistic with  decay parameter $tau = 0.9$ (`gwesp.fixed.0.9'). We again assume a common group-level covariance structure and focus on the group-level mean parameters. The full model is given by:
\begin{equation}
\begin{split}
\pmb{y}^{(i)} &\sim p(\cdot | \theta^{(i)}), ~i = 1, \dots, 200, \\
\theta^{(i)} &\sim \mathcal{N}(\mu^{(g_i)}, \Sigma_\theta), ~i = 1, \dots, 200, \\ 
\mu^{(j)} &\sim \mathcal{N}(\mu^{pop}, \pmb{\Sigma}_\mu), ~j = 1, 2 \\
\pmb{\Sigma}_\theta &\sim \mathcal{W}^{-1}(\nu_\theta, \pmb{\Psi}_\theta) \\
\mu^{pop} &\sim \mathcal{N}(\mu_0, \pmb{\Lambda}) \\
\pmb{\Sigma}_\mu &\sim \mathcal{W}^{-1}(\nu_\mu,\pmb{\Psi}_\mu).
\end{split}
\end{equation}

\begin{figure}
    \centering
    \includegraphics[width=\textwidth]{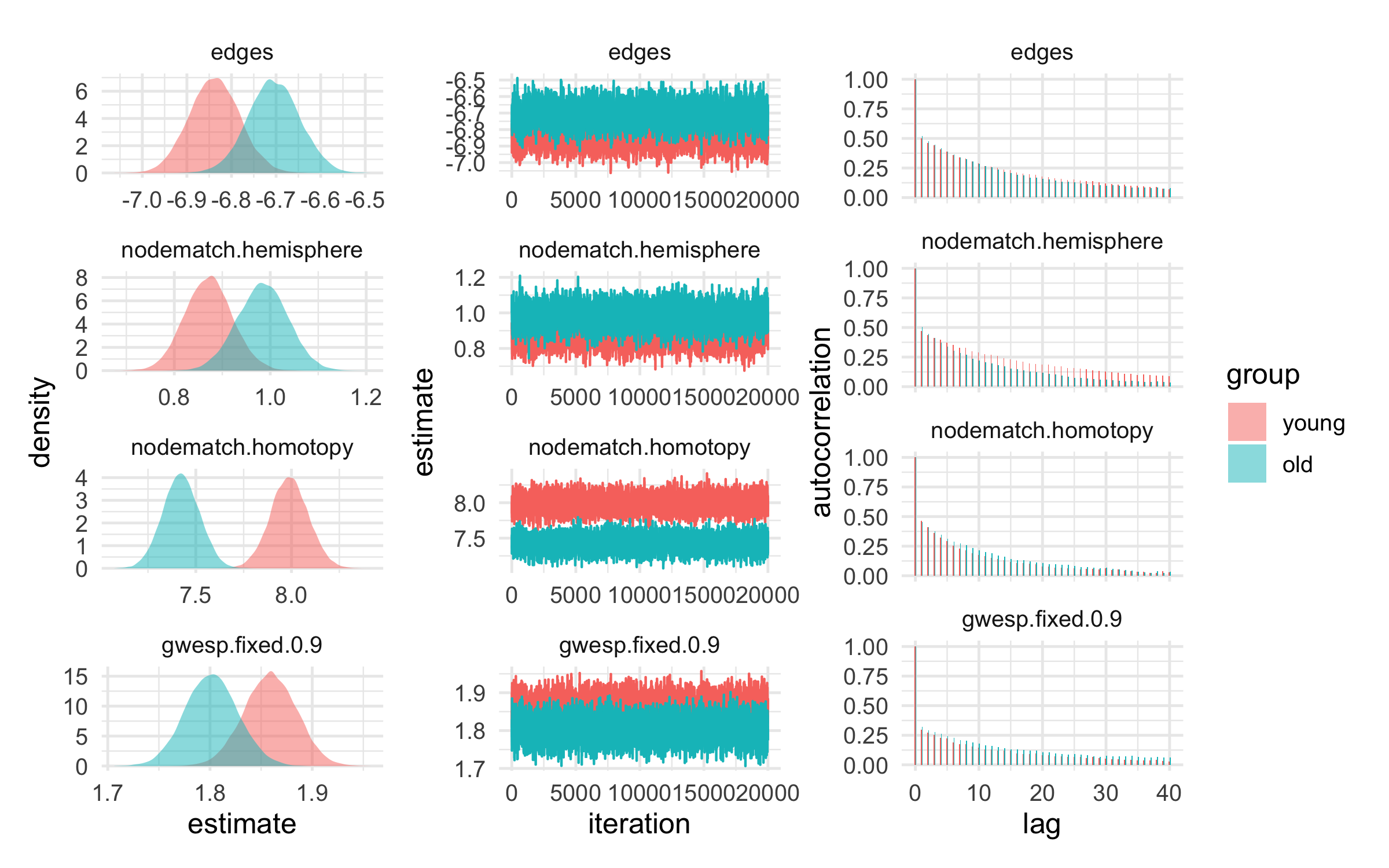}
    \caption{MCMC output from the exchange-within-Gibbs algorithm for the group-level mean parameters of a population of resting-state fMRI networks from a group of 100 young individuals and a group of 100 old individuals. }
    \label{fig:brain}
\end{figure}

We used the exchange-within-Gibbs algorithm with ASIS to generate 22,000 posterior samples, discarding the first 2,000 samples as burn-in. Figure~\ref{fig:brain} shows summaries of the posterior samples for $(\mu^{(1)}, \mu^{(2)})$, with the trace and autocorrelation plots demonstrating that the MCMC has mixed well. The posterior density plots show that the clearest difference between the old group and the young group was the difference in the parameter associated with the number of edges between homotomic nodes (`nodematch.homotopy'). While this parameter is large and positive for both groups, it is moderately smaller in the old group, indicating that the propensity for homotopic connections is lower in old age. On the other hand, there is no clear evidence for group differences in the remaining parameters. The edges parameters are large and negative, pointing to the overal sparsity of the networks; the intrahemisphere parameters (`nodematch.hemisphere') are small and positive, indicating a moderate propensity for connections between nodes in the same half of the brain; and the GWESP parameters are also positive, indicating a propensity to form triangles and thus a degree of functional segregation \citep{Bullmore2009}. 

\begin{figure}
    \centering
    \includegraphics[width=\textwidth]{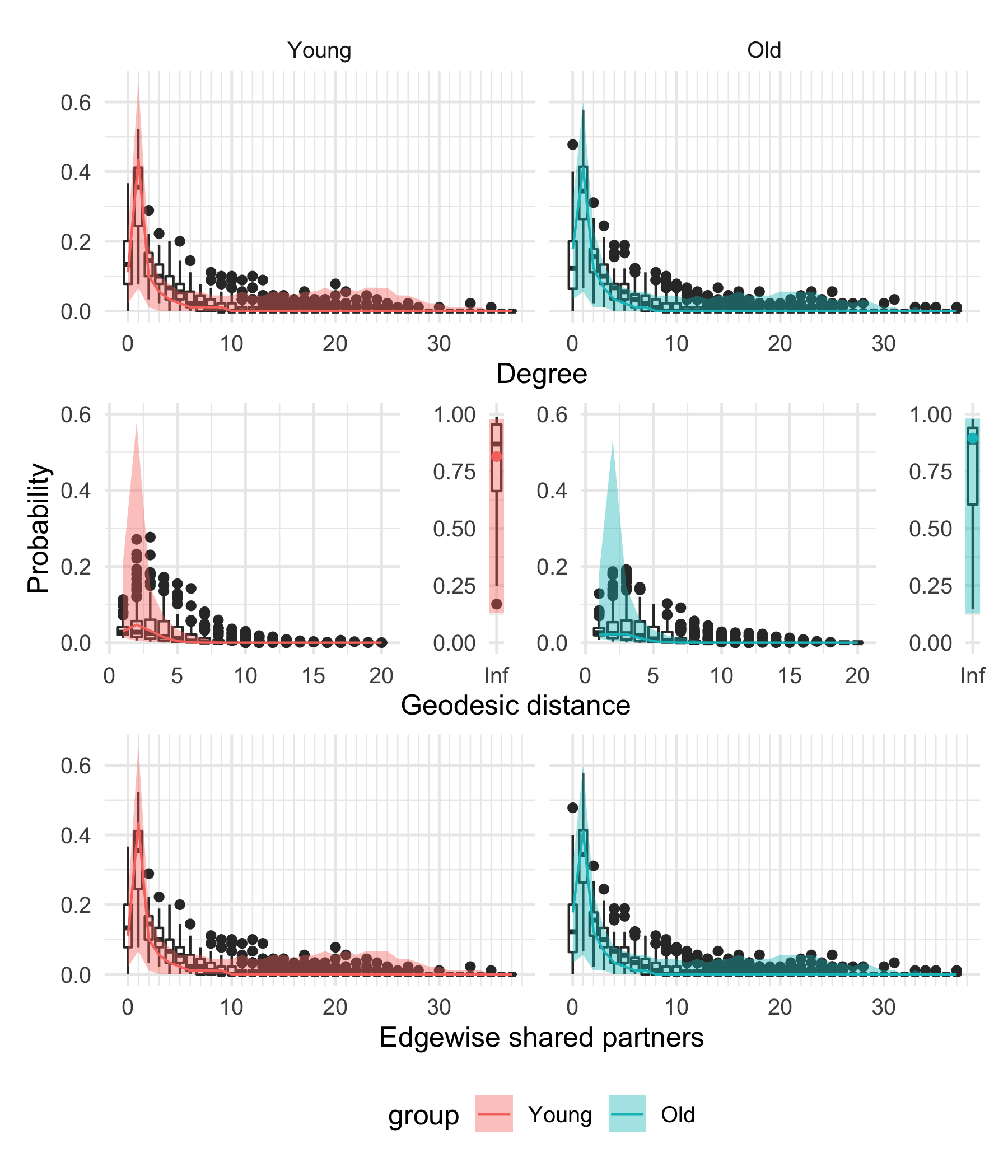}
    \caption{Graphical goodness-of-fit assessment for resting-state fMRI networks from a young group and an old group, fitted in a joint model. The box plots correspond to the observed networks, while the ribbons represent 95\% credible intervals corresponding to the posterior predictive networks. Note that a geodesic distance of infinity between two nodes means that there is no path connecting the nodes.}
    \label{fig:brain_gof}
\end{figure}

To assess goodness-of-fit, for both groups we generated $S=100$ networks from the model at posterior samples of $\mu^{(j)}$ chosen uniformly at random. Figure \ref{fig:brain_gof} indicates a reasonable fit for both groups, with the geodesic distance and edgewise shared partner distributions showing a good correspondence between the simulated networks and the observed networks. There appears to be a slight discrepancy in the degree distributions, with the simulated networks in the young group in particular having fewer nodes of degree 4 to 6 relative to the observed networks. 

\section{Discussion}

The main contribution of this article is to introduce a multilevel framework for modelling populations of networks, along with a novel MCMC procedure for performing inference with the framework. While the framework itself is a natural multilevel extension of a single ERGM, the inference procedure is more involved due to the intractability of the ERGM likelihood and the challenges associated with MCMC for hierarchical models. We have presented how our framework can be applied to functional connectivity networks from two groups of individuals. The framework allows the pooling of information across individuals within the same group and, depending on the parametrisation, across groups. By analysing the posterior samples produced by the MCMC algorithm, we can determine whether there are statistically significant differences in the connectivity structure between the two groups. Although we chose here to focus on networks constructed from resting-state fMRI scans, our framework could also be applied to networks derived from other neuroimaging modalities such as magnetoencephalography (MEG) or diffusion tensor imaging (DTI). 

The flexibility of the multilevel framework could be exploited to model more complex group structures such as multiple groups or factorial designs. Similarly, given multiple network measurements per individual, it is straightforward (conceptually, if not computationally) to extend the framework by adding another layer to the model. One could also incorporate individual-level covariate information into the model. In the context of the healthy ageing example considered above, this would avoid splitting individuals into (somewhat arbitrary) groups and provide a mechanism to measure how connectivity structure varies throughout the lifespan.

Another important extension would be to use weighted exponential random graph models \citep{Krivitsky2012, Desmarais2012}. These are an extension of the binary ERGM that can be applied to weighted networks, thus avoiding the thresholding step in the construction of functional connectivity networks. Indeed, one version of a weighted ERGM, the generalised exponential random graph model (GERGM) \citep{Desmarais2012} was recently applied to a 20-node functional connectivity network \citep{Stillman2017}. This approach has the additional advantage of modelling the mean connectivity directly and thus would avoid any confounding due to differences in mean connectivity. However, the GERGM is at present extremely computational intensive, rendering it infeasible for a population of networks.

One of the key challenges in applying our framework to real data is the choice of which network summary statistics to include in the model. A fully Bayesian model selection method based on reversible-jump MCMC has been developed for exponential random graph models on single networks \citep{Caimo2013}. A similar approach could be developed for our framework, though the computational cost is likely to be prohibitive. A more pragmatic approach would be to develop a graphical goodness-of-fit method by comparing the posterior predictive distributions under different models.

The computational cost of our MCMC algorithm is considerable. Using a 20-core computing cluster (Intel Skylake, 2.6 GHz), the algorithm took 29 hours to produce the 22,000 posterior samples in the real data example presented above. The main computational bottleneck lies in simulating the exponential random graphs at each MCMC iteration. While the computational cost should increase roughly linearly in the number of networks, \cite{Krivitsky2014supp} provide empirical evidence indicating that the cost may grow on the order of $p(N + E)\log(E)$ where $p$ is the number of summary statistics, $N$ is the number of nodes, and $E$ is the number of edges. It may be possible to reduce the number of ERGM simulations at each MCMC iteration using noisy Monte Carlo methods \citep{Alquier2016}. Another promising avenue is variational inference for ERGMs \citep{tan2020}, which could be extended to our framework to yield approximate Bayesian inference at a much reduced computational cost relative to MCMC.

%An alternative approach would be to explicitly account for noise in the model. Recently, a method \cite{Peixoto2018} was proposed to reconstruct networks in the presence of noisy data by specifying the generative model for the network jointly with the noise measurement process. While this approach focused on the stochastic block model, the idea should still be applicable to ERGMs. Since higher levels of noise generally result in lower mean connectivity, accounting for the noise should lead to more reliable estimates of the overall connectivity structure. 

\section*{Acknowledgements}

B.L. and S.W. were supported by the UK Medical Research Council [Programme number U105292687]. B.L. was also supported by the UK Engineering and Physical Sciences Research Council through the Bayes4Health programme [Grant number EP/R018561/1] and gratefully acknowledges funding from Jesus College, Oxford. This research was supported by the NIHR Cambridge Biomedical Research Centre (BRC-1215-20014). The computational aspects of this research were supported by the Wellcome Trust Core Award Grant Number 203141/Z/16/Z and the NIHR Oxford BRC. The views expressed are those of the authors and not necessarily those of the NHS, the NIHR or the Department of Health and Social Care.

\bibliographystyle{ba}
\bibliography{multibergm}

\begin{thebibliography}{58}
\newcommand{\enquote}[1]{``#1''}
\expandafter\ifx\csname natexlab\endcsname\relax\def\natexlab#1{#1}\fi
\expandafter\ifx\csname url\endcsname\relax
  \def\url#1{{\tt #1}}\fi
\expandafter\ifx\csname urlprefix\endcsname\relax\def\urlprefix{URL }\fi
\ifx\endbibitem\undefined \let\endbibitem\relax\fi

\bibitem[{Achard et~al.(2006)Achard, Salvador, Whitcher, Suckling, and
  Bullmore}]{Achard2006}
Achard, S., Salvador, R., Whitcher, B., Suckling, J., and Bullmore, E. (2006).
\newblock \enquote{A Resilient, Low-Frequency, Small-World Human Brain
  Functional Network with Highly Connected Association Cortical Hubs.}
\newblock {\em Journal of Neuroscience\/}, 26(1): 63--72.
\newline\urlprefix\url{http://www.jneurosci.org/content/26/1/63}
\endbibitem

\bibitem[{Alquier et~al.(2016)Alquier, Friel, Everitt, and
  Boland}]{Alquier2016}
Alquier, P., Friel, N., Everitt, R., and Boland, A. (2016).
\newblock \enquote{Noisy Monte Carlo: convergence of Markov chains with
  approximate transition kernels.}
\newblock {\em Statistics and Computing\/}, 26(1): 29--47.
\newline\urlprefix\url{https://doi.org/10.1007/s11222-014-9521-x}
\endbibitem

\bibitem[{Bullmore and Sporns(2009)}]{Bullmore2009}
Bullmore, E. and Sporns, O. (2009).
\newblock \enquote{Complex brain networks: graph theoretical analysis of
  structural and functional systems.}
\newblock {\em Nat Rev Neurosci\/}, 10(3): 186--198.
\newline\urlprefix\url{http://dx.doi.org/10.1038/nrn2575}
\endbibitem

\bibitem[{Butts(2018)}]{Butts2018}
Butts, C.~T. (2018).
\newblock \enquote{A perfect sampling method for exponential family random
  graph models.}
\newblock {\em The Journal of Mathematical Sociology\/}, 42(1): 17--36.
\endbibitem

\bibitem[{Caimo and Friel(2011)}]{Caimo2011}
Caimo, A. and Friel, N. (2011).
\newblock \enquote{Bayesian inference for exponential random graph models.}
\newblock {\em Social Networks\/}, 33(1): 41 -- 55.
\newline\urlprefix\url{http://www.sciencedirect.com/science/article/pii/S0378873310000493}
\endbibitem

\bibitem[{Caimo and Friel(2013)}]{Caimo2013}
--- (2013).
\newblock \enquote{Bayesian model selection for exponential random graph
  models.}
\newblock {\em Social Networks\/}, 35(1): 11 -- 24.
\newline\urlprefix\url{http://www.sciencedirect.com/science/article/pii/S0378873312000573}
\endbibitem

\bibitem[{Desmarais and Cranmer(2012)}]{Desmarais2012}
Desmarais, B.~A. and Cranmer, S.~J. (2012).
\newblock \enquote{Statistical inference for valued-edge networks: The
  generalized exponential random graph model.}
\newblock {\em PloS one\/}, 7(1): e30136.
\endbibitem

\bibitem[{Durante et~al.(2017)Durante, Dunson, and Vogelstein}]{Durante2017}
Durante, D., Dunson, D.~B., and Vogelstein, J.~T. (2017).
\newblock \enquote{Nonparametric Bayes modeling of populations of networks.}
\newblock {\em Journal of the American Statistical Association\/}, 112(520):
  1516--1530.
\endbibitem

\bibitem[{Durante et~al.(2018)Durante, Dunson et~al.}]{Durante2018}
Durante, D., Dunson, D.~B., et~al. (2018).
\newblock \enquote{Bayesian inference and testing of group differences in brain
  networks.}
\newblock {\em Bayesian Analysis\/}, 13(1): 29--58.
\endbibitem

\bibitem[{Everitt(2012)}]{Everitt2012}
Everitt, R.~G. (2012).
\newblock \enquote{Bayesian Parameter Estimation for Latent Markov Random
  Fields and Social Networks.}
\newblock {\em Journal of Computational and Graphical Statistics\/}, 21(4):
  940--960.
\newline\urlprefix\url{https://doi.org/10.1080/10618600.2012.687493}
\endbibitem

\bibitem[{Fallani et~al.(2017)Fallani, Latora, and Chavez}]{Fallani2017}
Fallani, F. D.~V., Latora, V., and Chavez, M. (2017).
\newblock \enquote{A {Topological} {Criterion} for {Filtering} {Information} in
  {Complex} {Brain} {Networks}.}
\newblock {\em PLOS Computational Biology\/}, 13(1): e1005305.
\newline\urlprefix\url{https://journals.plos.org/ploscompbiol/article?id=10.1371/journal.pcbi.1005305}
\endbibitem

\bibitem[{Frank and Strauss(1986)}]{Frank1986}
Frank, O. and Strauss, D. (1986).
\newblock \enquote{Markov graphs.}
\newblock {\em Journal of the american Statistical association\/}, 81(395):
  832--842.
\endbibitem

\bibitem[{Fuster(2006)}]{Fuster2006}
Fuster, J.~M. (2006).
\newblock \enquote{The cognit: {A} network model of cortical representation.}
\newblock {\em International Journal of Psychophysiology\/}, 60(2): 125 -- 132.
\newline\urlprefix\url{http://www.sciencedirect.com/science/article/pii/S0167876006000079}
\endbibitem

\bibitem[{Geerligs et~al.(2017)Geerligs, Tsvetanov, Cam-CAN, and
  Henson}]{Geerligs2017}
Geerligs, L., Tsvetanov, K.~A., Cam-CAN, and Henson, R.~N. (2017).
\newblock \enquote{Challenges in measuring individual differences in functional
  connectivity using fMRI: The case of healthy aging.}
\newblock {\em Human Brain Mapping\/}, 38(8): 4125--4156.
\newline\urlprefix\url{https://onlinelibrary.wiley.com/doi/abs/10.1002/hbm.23653}
\endbibitem

\bibitem[{Gelfand et~al.(1995)Gelfand, Sahu, and Carlin}]{Gelfand1995}
Gelfand, A.~E., Sahu, S.~K., and Carlin, B.~P. (1995).
\newblock \enquote{Efficient Parametrisations for Normal Linear Mixed Models.}
\newblock {\em Biometrika\/}, 82(3): 479--488.
\newline\urlprefix\url{http://www.jstor.org/stable/2337527}
\endbibitem

\bibitem[{Geman and Geman(1984)}]{Geman1984}
Geman, S. and Geman, D. (1984).
\newblock \enquote{Stochastic Relaxation, Gibbs Distributions, and the Bayesian
  Restoration of Images.}
\newblock {\em IEEE Transactions on Pattern Analysis and Machine
  Intelligence\/}, PAMI-6(6): 721--741.
\endbibitem

\bibitem[{Ginestet et~al.(2017)Ginestet, Li, Balachandran, Rosenberg, and
  Kolaczyk}]{Ginestet2017}
Ginestet, C.~E., Li, J., Balachandran, P., Rosenberg, S., and Kolaczyk, E.~D.
  (2017).
\newblock \enquote{Hypothesis testing for network data in functional
  neuroimaging.}
\newblock {\em Ann. Appl. Stat.\/}, 11(2): 725--750.
\newline\urlprefix\url{https://doi.org/10.1214/16-AOAS1015}
\endbibitem

\bibitem[{Ginestet et~al.(2011)Ginestet, Nichols, Bullmore, and
  Simmons}]{Ginestet2011}
Ginestet, C.~E., Nichols, T.~E., Bullmore, E.~T., and Simmons, A. (2011).
\newblock \enquote{Brain Network Analysis: Separating Cost from Topology Using
  Cost-Integration.}
\newblock {\em PLOS ONE\/}, 6(7): 1--17.
\newline\urlprefix\url{https://doi.org/10.1371/journal.pone.0021570}
\endbibitem

\bibitem[{Haario et~al.(2001)Haario, Saksman, Tamminen
  et~al.}]{haario2001adaptive}
Haario, H., Saksman, E., Tamminen, J., et~al. (2001).
\newblock \enquote{An adaptive Metropolis algorithm.}
\newblock {\em Bernoulli\/}, 7(2): 223--242.
\endbibitem

\bibitem[{Hastings(1970)}]{Hastings1970}
Hastings, W.~K. (1970).
\newblock \enquote{Monte Carlo Sampling Methods Using Markov Chains and Their
  Applications.}
\newblock {\em Biometrika\/}, 57(1): 97--109.
\newline\urlprefix\url{http://www.jstor.org/stable/2334940}
\endbibitem

\bibitem[{Hoff et~al.(2002)Hoff, Raftery, and Handcock}]{Hoff2002}
Hoff, P.~D., Raftery, A.~E., and Handcock, M.~S. (2002).
\newblock \enquote{Latent space approaches to social network analysis.}
\newblock {\em Journal of the american Statistical association\/}, 97(460):
  1090--1098.
\endbibitem

\bibitem[{Holland et~al.(1983)Holland, Laskey, and Leinhardt}]{Holland1983}
Holland, P.~W., Laskey, K.~B., and Leinhardt, S. (1983).
\newblock \enquote{Stochastic blockmodels: First steps.}
\newblock {\em Social networks\/}, 5(2): 109--137.
\endbibitem

\bibitem[{Hunter et~al.(2008{\natexlab{a}})Hunter, Handcock, Butts, Goodreau,
  and Morris}]{Hunter2008a}
Hunter, D., Handcock, M., Butts, C., Goodreau, S., and Morris, M.
  (2008{\natexlab{a}}).
\newblock \enquote{ergm: A Package to Fit, Simulate and Diagnose
  Exponential-Family Models for Networks.}
\newblock {\em Journal of Statistical Software, Articles\/}, 24(3): 1--29.
\newline\urlprefix\url{https://www.jstatsoft.org/v024/i03}
\endbibitem

\bibitem[{Hunter et~al.(2008{\natexlab{b}})Hunter, Goodreau, and
  Handcock}]{Hunter2008b}
Hunter, D.~R., Goodreau, S.~M., and Handcock, M.~S. (2008{\natexlab{b}}).
\newblock \enquote{Goodness of Fit of Social Network Models.}
\newblock {\em Journal of the American Statistical Association\/}, 103(481):
  248--258.
\newline\urlprefix\url{https://doi.org/10.1198/016214507000000446}
\endbibitem

\bibitem[{Kolaczyk(2009)}]{Kolaczyk2009}
Kolaczyk, E.~D. (2009).
\newblock {\em Statistical {Analysis} of {Network} {Data}: {Methods} and
  {Models}\/}.
\newblock Springer {Series} in {Statistics}. New York: Springer-Verlag.
\newline\urlprefix\url{//www.springer.com/gb/book/9780387881454}
\endbibitem

\bibitem[{Koskinen(2004)}]{Koskinen2004}
Koskinen, J. (2004).
\newblock \enquote{Bayesian analysis of exponential random graphs-estimation of
  parameters and model selection.}
\newblock Technical report, Stockholm University.
\endbibitem

\bibitem[{Koskinen et~al.(2013)Koskinen, Robins, Wang, and
  Pattison}]{Koskinen2013}
Koskinen, J.~H., Robins, G.~L., Wang, P., and Pattison, P.~E. (2013).
\newblock \enquote{Bayesian analysis for partially observed network data,
  missing ties, attributes and actors.}
\newblock {\em Social Networks\/}, 35(4): 514--527.
\newline\urlprefix\url{http://www.sciencedirect.com/science/article/pii/S0378873313000671}
\endbibitem

\bibitem[{Krivitsky(2012)}]{Krivitsky2012}
Krivitsky, P.~N. (2012).
\newblock \enquote{Exponential-family random graph models for valued networks.}
\newblock {\em Electron. J. Statist.\/}, 6: 1100--1128.
\newline\urlprefix\url{http://dx.doi.org/10.1214/12-EJS696}
\endbibitem

\bibitem[{Krivitsky and Handcock(2014)}]{Krivitsky2014supp}
Krivitsky, P.~N. and Handcock, M.~S. (2014).
\newblock \enquote{Supplementary Material: A separable model for dynamic
  networks.}
\newblock {\em Journal of the Royal Statistical Society: Series B (Statistical
  Methodology)\/}, 76(1): 29--46.
\newline\urlprefix\url{https://rss.onlinelibrary.wiley.com/doi/abs/10.1111/rssb.12014}
\endbibitem

\bibitem[{Lehmann et~al.(2021)Lehmann, Henson, Geerligs, Cam-CAN, and
  White}]{lehmann2021}
Lehmann, B., Henson, R., Geerligs, L., Cam-CAN, and White, S. (2021).
\newblock \enquote{Characterising group-level brain connectivity: A framework
  using Bayesian exponential random graph models.}
\newblock {\em NeuroImage\/}, 225: 117480.
\newline\urlprefix\url{https://www.sciencedirect.com/science/article/pii/S1053811920309654}
\endbibitem

\bibitem[{Mukherjee et~al.(2017)Mukherjee, Sarkar, and Lin}]{mukherjee2017}
Mukherjee, S.~S., Sarkar, P., and Lin, L. (2017).
\newblock \enquote{On Clustering Network-Valued Data.}
\newblock In {\em Proceedings of the 31st International Conference on Neural
  Information Processing Systems\/}, NIPS'17, 7074–7084. Red Hook, NY, USA:
  Curran Associates Inc.
\endbibitem

\bibitem[{Murray et~al.(2006)Murray, Ghahramani, and MacKay}]{Murray2006}
Murray, I., Ghahramani, Z., and MacKay, D. J.~C. (2006).
\newblock \enquote{{MCMC} for doubly-intractable distributions.}
\newblock In {\em Proceedings of the 22nd Annual Conference on Uncertainty in
  Artificial Intelligence (UAI-06)\/}, 359--366. AUAI Press.
\endbibitem

\bibitem[{Obando and Fallani(2017)}]{Obando2017}
Obando, C. and Fallani, F. D.~V. (2017).
\newblock \enquote{A statistical model for brain networks inferred from
  large-scale electrophysiological signals.}
\newblock {\em Journal of The Royal Society Interface\/}, 14(128): 20160940.
\newline\urlprefix\url{http://rsif.royalsocietypublishing.org/content/14/128/20160940}
\endbibitem

\bibitem[{Papaspiliopoulos et~al.(2003)Papaspiliopoulos, Roberts, and
  Sköld}]{Papaspiliopoulos2003}
Papaspiliopoulos, O., Roberts, G.~O., and Sköld, M. (2003).
\newblock \enquote{Non-centered parameterisations for hierarchical models and
  data augmentation.}
\newblock In Bernardo, J., Bayarri, M., Berger, J., Dawid, A., Heckerman, D.,
  Smith, A., and West, M. (eds.), {\em Bayesian Statistics 7: Proceedings of
  the Seventh Valencia International Meeting\/}, volume 307. Oxford University
  Press, USA.
\endbibitem

\bibitem[{Papaspiliopoulos et~al.(2007)Papaspiliopoulos, Roberts, and
  Sköld}]{Papaspiliopoulos2007}
--- (2007).
\newblock \enquote{A General Framework for the Parametrization of Hierarchical
  Models.}
\newblock {\em Statist. Sci.\/}, 22(1): 59--73.
\newline\urlprefix\url{https://doi.org/10.1214/088342307000000014}
\endbibitem

\bibitem[{Park and Haran(2018)}]{Park2018}
Park, J. and Haran, M. (2018).
\newblock \enquote{Bayesian inference in the presence of intractable
  normalizing functions.}
\newblock {\em Journal of the American Statistical Association\/}, 113(523):
  1372--1390.
\endbibitem

\bibitem[{Roberts et~al.(1997)Roberts, Gelman, Gilks et~al.}]{roberts1997weak}
Roberts, G.~O., Gelman, A., Gilks, W.~R., et~al. (1997).
\newblock \enquote{Weak convergence and optimal scaling of random walk
  Metropolis algorithms.}
\newblock {\em Annals of Applied probability\/}, 7(1): 110--120.
\endbibitem

\bibitem[{Roberts and Rosenthal(2009)}]{roberts2009examples}
Roberts, G.~O. and Rosenthal, J.~S. (2009).
\newblock \enquote{Examples of adaptive MCMC.}
\newblock {\em Journal of Computational and Graphical Statistics\/}, 18(2):
  349--367.
\endbibitem

\bibitem[{Roberts et~al.(2001)Roberts, Rosenthal et~al.}]{roberts2001optimal}
Roberts, G.~O., Rosenthal, J.~S., et~al. (2001).
\newblock \enquote{Optimal scaling for various Metropolis-Hastings algorithms.}
\newblock {\em Statistical science\/}, 16(4): 351--367.
\endbibitem

\bibitem[{Schweinberger and Handcock(2015)}]{Schweinberger2015}
Schweinberger, M. and Handcock, M.~S. (2015).
\newblock \enquote{Local dependence in random graph models: characterization,
  properties and statistical inference.}
\newblock {\em Journal of the Royal Statistical Society: Series B (Statistical
  Methodology)\/}, 77(3): 647--676.
\newline\urlprefix\url{https://rss.onlinelibrary.wiley.com/doi/abs/10.1111/rssb.12081}
\endbibitem

\bibitem[{Schweinberger et~al.(2020)Schweinberger, Krivitsky, Butts, and
  Stewart}]{Schweinberger2020}
Schweinberger, M., Krivitsky, P.~N., Butts, C.~T., and Stewart, J.~R. (2020).
\newblock \enquote{Exponential-Family Models of Random Graphs: Inference in
  Finite, Super and Infinite Population Scenarios.}
\newblock {\em Statist. Sci.\/}, 35(4): 627--662.
\newline\urlprefix\url{https://doi.org/10.1214/19-STS743}
\endbibitem

\bibitem[{Shafto et~al.(2014)Shafto, Tyler, Dixon, Taylor, Rowe, Cusack,
  Calder, Marslen-Wilson, Duncan, Dalgleish, Henson, Brayne, and
  Matthews}]{Shafto2014}
Shafto, M.~A., Tyler, L.~K., Dixon, M., Taylor, J.~R., Rowe, J.~B., Cusack, R.,
  Calder, A.~J., Marslen-Wilson, W.~D., Duncan, J., Dalgleish, T., Henson,
  R.~N., Brayne, C., and Matthews, F.~E. (2014).
\newblock \enquote{The Cambridge Centre for Ageing and Neuroscience (Cam-CAN)
  study protocol: a cross-sectional, lifespan, multidisciplinary examination of
  healthy cognitive ageing.}
\newblock {\em BMC Neurology\/}, 14(1): 1--25.
\newline\urlprefix\url{http://dx.doi.org/10.1186/s12883-014-0204-1}
\endbibitem

\bibitem[{Signorelli and Wit(2020)}]{signorelli2020}
Signorelli, M. and Wit, E.~C. (2020).
\newblock \enquote{Model-based clustering for populations of networks.}
\newblock {\em Statistical Modelling\/}, 20(1): 9--29.
\endbibitem

\bibitem[{Simpson et~al.(2013)Simpson, Lyday, Hayasaka, Marsh, and
  Laurienti}]{Simpson2013}
Simpson, S., Lyday, R., Hayasaka, S., Marsh, A., and Laurienti, P. (2013).
\newblock \enquote{A permutation testing framework to compare groups of brain
  networks.}
\newblock {\em Frontiers in Computational Neuroscience\/}, 7: 171.
\newline\urlprefix\url{https://www.frontiersin.org/article/10.3389/fncom.2013.00171}
\endbibitem

\bibitem[{Simpson et~al.(2011)Simpson, Hayasaka, and Laurienti}]{Simpson2011}
Simpson, S.~L., Hayasaka, S., and Laurienti, P.~J. (2011).
\newblock \enquote{Exponential random graph modeling for complex brain
  networks.}
\newblock {\em PloS one\/}, 6(5): e20039.
\endbibitem

\bibitem[{Sinke et~al.(2016)Sinke, Dijkhuizen, Caimo, Stam, and
  Otte}]{Sinke2016}
Sinke, M.~R., Dijkhuizen, R.~M., Caimo, A., Stam, C.~J., and Otte, W.~M.
  (2016).
\newblock \enquote{Bayesian exponential random graph modeling of whole-brain
  structural networks across lifespan.}
\newblock {\em NeuroImage\/}, 135(Supplement C): 79 -- 91.
\newline\urlprefix\url{http://www.sciencedirect.com/science/article/pii/S1053811916301069}
\endbibitem

\bibitem[{Slaughter and Koehly(2016)}]{slaughter2016}
Slaughter, A.~J. and Koehly, L.~M. (2016).
\newblock \enquote{Multilevel models for social networks: hierarchical Bayesian
  approaches to exponential random graph modeling.}
\newblock {\em Social networks\/}, 44: 334--345.
\endbibitem

\bibitem[{Stillman et~al.(2017)Stillman, Wilson, Denny, Desmarais, Bhamidi,
  Cranmer, and Lu}]{Stillman2017}
Stillman, P.~E., Wilson, J.~D., Denny, M.~J., Desmarais, B.~A., Bhamidi, S.,
  Cranmer, S.~J., and Lu, Z.-L. (2017).
\newblock \enquote{Statistical Modeling of the Default Mode Brain Network
  Reveals a Segregated Highway Structure.}
\newblock {\em Scientific Reports\/}, 7(1): 11694.
\newline\urlprefix\url{https://doi.org/10.1038/s41598-017-09896-6}
\endbibitem

\bibitem[{Sweet et~al.(2013)Sweet, Thomas, and Junker}]{Sweet2013}
Sweet, T.~M., Thomas, A.~C., and Junker, B.~W. (2013).
\newblock \enquote{Hierarchical network models for education research:
  Hierarchical latent space models.}
\newblock {\em Journal of Educational and Behavioral Statistics\/}, 38(3):
  295--318.
\endbibitem

\bibitem[{Sweet et~al.(2014)Sweet, Thomas, and Junker}]{Sweet2014}
--- (2014).
\newblock \enquote{Hierarchical mixed membership stochastic blockmodels for
  multiple networks and experimental interventions.}
\newblock In Airoldi, E.~M., Blei, D., Erosheva, E.~A., and Fienberg, S.~E.
  (eds.), {\em Handbook of Mixed Membership Models and Their Applications\/},
  463--488. Chapman \& Hall/CRC, 1st edition.
\endbibitem

\bibitem[{Tan and Friel(2020)}]{tan2020}
Tan, L.~S. and Friel, N. (2020).
\newblock \enquote{Bayesian variational inference for exponential random graph
  models.}
\newblock {\em Journal of Computational and Graphical Statistics\/}, 1--19.
\endbibitem

\bibitem[{Thiemichen et~al.(2016)Thiemichen, Friel, Caimo, and
  Kauermann}]{Thiemichen2016}
Thiemichen, S., Friel, N., Caimo, A., and Kauermann, G. (2016).
\newblock \enquote{Bayesian exponential random graph models with nodal random
  effects.}
\newblock {\em Social Networks\/}, 46: 11--28.
\newline\urlprefix\url{http://www.sciencedirect.com/science/article/pii/S0378873316000034}
\endbibitem

\bibitem[{Tierney(1994)}]{Tierney1994}
Tierney, L. (1994).
\newblock \enquote{Markov Chains for Exploring Posterior Distributions.}
\newblock {\em Ann. Statist.\/}, 22(4): 1701--1728.
\newline\urlprefix\url{https://doi.org/10.1214/aos/1176325750}
\endbibitem

\bibitem[{Tzourio-Mazoyer et~al.(2002)Tzourio-Mazoyer, Landeau, Papathanassiou,
  Crivello, Etard, Delcroix, Mazoyer, and Joliot}]{TzourioMazoyer2002}
Tzourio-Mazoyer, N., Landeau, B., Papathanassiou, D., Crivello, F., Etard, O.,
  Delcroix, N., Mazoyer, B., and Joliot, M. (2002).
\newblock \enquote{Automated Anatomical Labeling of Activations in SPM Using a
  Macroscopic Anatomical Parcellation of the MNI MRI Single-Subject Brain.}
\newblock {\em NeuroImage\/}, 15(1): 273 -- 289.
\newline\urlprefix\url{http://www.sciencedirect.com/science/article/pii/S1053811901909784}
\endbibitem

\bibitem[{Wang and Atchad{\'e}(2014)}]{wang2014}
Wang, J. and Atchad{\'e}, Y.~F. (2014).
\newblock \enquote{Approximate Bayesian computation for exponential random
  graph models for large social networks.}
\newblock {\em Communications in Statistics-Simulation and Computation\/},
  43(2): 359--377.
\endbibitem

\bibitem[{Wang et~al.(2013)Wang, Robins, Pattison, and Lazega}]{wang2013}
Wang, P., Robins, G., Pattison, P., and Lazega, E. (2013).
\newblock \enquote{Exponential random graph models for multilevel networks.}
\newblock {\em Social Networks\/}, 35(1): 96--115.
\endbibitem

\bibitem[{Yin et~al.(2020)Yin, Shen, and Butts}]{yin2020}
Yin, F., Shen, W., and Butts, C.~T. (2020).
\newblock \enquote{Finite Mixtures of ERGMs for Modeling Ensembles of
  Networks.}
\endbibitem

\bibitem[{Yu and Meng(2011)}]{Yu2011}
Yu, Y. and Meng, X.-L. (2011).
\newblock \enquote{To Center or Not to Center: That Is Not the Question—An
  Ancillarity–Sufficiency Interweaving Strategy (ASIS) for Boosting MCMC
  Efficiency.}
\newblock {\em Journal of Computational and Graphical Statistics\/}, 20(3):
  531--570.
\newline\urlprefix\url{https://doi.org/10.1198/jcgs.2011.203main}
\endbibitem

\end{thebibliography}

% \begin{figure} 
% \includegraphics{<eps-file>}% place <eps-file> in ./img  subfolder
% \caption{}
% \label{}
% \end{figure}

% \begin{table} 
% *****************
% \begin{tabular}{lll}
% \end{tabular}
% *****************
% \caption{}
% \label{}
% \end{figure}

%%%%%%%%%%%%%%%%%%%%%%%%%%%%%%%%%%%%%%%%%%%%%%
%% Supplementary Material, if any, should   %%
%% be provided in {supplement} environment  %%
%% with title and short description.        %%
%%%%%%%%%%%%%%%%%%%%%%%%%%%%%%%%%%%%%%%%%%%%%%
%\begin{supplement}
%\stitle{???}
%\sdescription{???.}
%\end{supplement}

%% ** The bibliograhy **
%\bibliographystyle{ba}
%\bibliography{<bib-data-file>}% place <bib-data-file> in ./bib folder 

% ** Acknowledgements **
% \begin{acknowledgement}
% \end{acknowledgement}

\end{document}